\documentclass[prb]{revtex4}
\usepackage{graphicx}

\def\figwidth{11cm}
\begin{document}
\title{Charge Density Waves and Bond Order Waves in a quarter filled
  extended Hubbard ladder} 
\author{E. Orignac}
\affiliation{Laboratoire de Physique Th\'eorique de l'\'Ecole Normale
  Sup\'erieure CNRS-UMR8549 \\ 24, Rue Lhomond \\ 75231 Paris Cedex 05
  France}
\author{R. Citro}
\affiliation{Dipartimento di Fisica ``E. R. Caianiello'' and
Unit{\`a} I.N.F.M. di Salerno\\ Universit{\`a} di Salerno, Via S.
Allende, 84081 Baronissi (Sa), Italy}

\date{\today}

\begin{abstract}
We investigate the phase diagram of a quarter filled Hubbard ladder
with nearest-neighbor Coulomb repulsion using bosonization and
renormalization group approach. Focusing on the strong-repulsion regime,
we discuss the effect of an interchain exchange interaction $J_\perp$
and interchain repulsion $V_\perp$ on the possible ground
states of the system and charge order configurations.
Since the spin excitations always possess a gap,
we find competing bond-order wave   and
charge density wave phases  as possible ground states of the
ladder model. We discuss the elementary excitations in these various
phases and point an analogy between the excitations on some of these
phases and those of a Kondo-Heisenberg insulator.
We also study the order of the quantum phase transitions
between the different ground states of the system. We obtain
second order transitions in the Ising or  $SU(2)_2$ universality class
or first order transitions.
We map the complete phase diagram in the $J_\perp-V_\perp$ plane
by integrating perturbative renormalization-group equations. Finally,
we discuss the effect of doping away from half-filling and the effect
of an applied magnetic field.
\end{abstract}

\maketitle

\section{Introduction}
Charge ordering is a general phenomenon in condensed matter physics
that has been recently observed in a variety of compounds
including rare
earth manganites\cite{cheong_mang,mori_mang},
quasi-one-dimensional $\mathrm{(TMTTF)_2X}$
conductors\cite{monceau_co_1d},  cuprate  or
nickelate materials\cite{tranquada_nick,tranquada_cup}. A very
particular form of charge order in these latter materials is the
formation of {\it charge stripes},
that are domain walls between hole rich and hole poor regions.
 From the theoretical point of view,
charge ordering is a crystallization of the electron liquid
which occurs when long-range Coulomb repulsion\cite{wigner} dominates
over kinetic energy. In one dimension, charge ordering and more
generally metal-insulator transitions can be studied in great details
due to the existence of powerful
analytical\cite{emery_revue_1d,solyom_revue_1d,schulz_houches_revue}
and numerical\cite{white_dmrg_letter} methods.
In the case of a one-dimensional system,
the simplest model\cite{hubbard_wigner}
of interacting electrons that allows for charge ordering is the
Hubbard model extended with an additional nearest-neighbor
Coulomb interaction $V$. Studies of the metal insulator transition in
one dimension show that charge ordered or Mott insulating states can
form at commensurate fillings for sufficiently strong
repulsion\cite{schulz_mott_revue,giamarchi_mott_shortrev}. In relation
with organic compounds, the quarter-filled extended Hubbard model on a
single chain has been studied in
details\cite{tsuchiizu_qf1d}. As a first step towards understanding
 charge ordering and stripe formation in strongly correlated electron models,
numerical studies of coupled chain systems at commensurate fillings
\cite{poilblanc_2ch,white_stripes_tj,bonca_3leg,tohyama_4leg,white_ladder_friedel} (ladders)
have also been performed. Half filled ladder systems
are Mott insulators that display a spin gap analogous to the Haldane
gap in spin-1 chains\cite{haldane_gap} when made of an even number of
coupled chains. Ladder models are not only of theoretical interest.
Quite a few half-filled ladder systems have been
synthesized both organic and inorganic and the spin gap behavior has
been characterized in great
details~\cite{takano_spingap,chaboussant_cuhpcl,iwase_cav2o5,rovira_dtttf,watson_bpcb,landee_ladder}.
Away from half-filling and more generally commensurate filling,
ladders are expected to be conducting. The persistence of the spin gap
away from half-filling has been proposed to give rise to a paring
mechanism and to superconductivity\cite{rice_srcuo}. This suggestion
has given rise to intense theoretical studies both
analytical\cite{fabrizio_2ch_rg,nagaosa_2ch,schulz_2chains,balents_2ch}
 and numerical
 \cite{fabrizio_q1d,poilblanc_2ch_mc,troyer_2ch_tj,hayward_2chain_2,noack_dmrg_2ch}
 that confirmed the presence of superconducting
 correlations at incommensurate fillings. Experimentally, the
 two-leg ladder compound $\mathrm{Sr_{14}Cu_{24}O_{41}}$\cite{uchara_SrCaCuO}
can be doped away from half filling and under pressure displays
a superconducting transition.
Recently, the compound $\alpha'-\mathrm{NaV_2O_5}$,
initially identified as an inorganic spin Peierls compound\cite{isobe_nav2o5}, was shown to be
a quarter-filled
ladder\cite{boer,smolinski_nav2o5,mostovoy_nav2o5,seo_nav2o5}, that
undergoes a  transition at $T_c=34K$
corresponding to the formation of a
charge ordered state with a zig-zag charge
pattern and the opening of a spin gap\cite{ohama_nav2o5,fagot-revurat_nav2o5,grenier_nav2o5,sawa_nav2o5}.
This has prompted numerical studies of
quarter-filled extended Hubbard two-leg ladder
systems\cite{vojta_qfl_short,vojta_qfl} that exhibit the
formation of a zig-zag charge ordered state at large repulsion.
Interestingly, the charge ordered state of the two-leg ladder at
quarter filling presents a spin gap in contrast to the single chain.
Analytic studies of insulating states on a two-leg ladder have been mostly
confined to the half-filled
case\cite{lin_so8,konik_ff_so8,wu_2leg_firstorder,tsuchiizu_2leg_firstorder}
or to mean field\cite{seo_nav2o5,amasaki_2ch_zigzag} approximations
and strong coupling
expansions\cite{mostovoy_nav2o5,horsch_nav2o5,sa_nav2o5} in the
quarter-filled case.
In the present paper we analyze the phase diagram of the quarter-filled two-leg
Hubbard ladder with interchain Coulomb repulsion $V_\perp$ and exchange
coupling $J_\perp$,
by using bosonization and renormalization group (RG) methods.
Focusing on the regime of strong repulsion, we discuss charge orders
and the elementary excitations of the insulating phase. Due to the
presence of a spin gap, the competing ground states are bond-order
wave states (BOW) and charge density wave states (CDW). We analyze the
phase transitions between these states using a perturbative
renormalization group approach and a mapping of spin excitations
to a Majorana fermion theory.  The plan of
the paper is the following. In Sec.~\ref{sec:hamiltonian} we introduce
the model and discuss the physics of two simple limits, the large
on-site repulsion and the large interchain exchange limit. In
Sec.~\ref{sec:bosonization} we give the details of the bosonization
treatment and of the derivation of the renormalization group (RG)
equations.
In Sec.~\ref{sec:phase} we introduce the order parameters of the
insulating phases and the describe the charge ordered ground states
obtained by the values of the phase fields that minimize the energy.
In Sec.~\ref{sec:excitations} we describe the elementary charge and
spin excitations of each ground state and discuss the analogy with
Kondo-Heisenberg-Hubbard chain in a particular case. In
Sec.~\ref{sec:numerics} we discuss the nature of the transitions among
the various phases in the antisymmetric charge sector and the spin
sector.  This is partially accomplished by a mapping to a theory of
Majorana fermions and by a mapping to an effective quantum Ising model
in the limit of strong interchain repulsion. Finally, we discuss the
results of the phase diagram obtained by the numerical integration of
the RG equations. We also briefly discuss the
commensurate-incommensurate transitions produced by the application
 of a magnetic field or by doping away from quarter-filling.

\section{Hamiltonian and some simple limits}
\label{sec:hamiltonian}

We consider the quarter-filled extended Hubbard model on a two leg
ladder. The Hamiltonian reads:
\begin{eqnarray}
  \label{eq:latt-hubbard}
  H&=&-t \sum_{i,p,\sigma} (c^\dagger_{i+1,p,\sigma}
  c_{i,p,\sigma}+c^\dagger_{i,p,\sigma} c_{i+1,p,\sigma}) -t_\perp
  \sum_{i,\sigma} (c^\dagger_{i,2,\sigma} c_{i,1,\sigma} +
  \text{H.c.}) \nonumber \\ && + U
  \sum_{i,p} n_{i,p,\uparrow} n_{i,p,\downarrow} + V_\parallel
  \sum_{i,p} n_{i,p} n_{i+1,p} + V_\perp \sum_i n_{i,1} n_{i,2},
\end{eqnarray}
\noindent where $n_{i,p,\sigma}=c^\dagger_{i,p,\sigma} c_{i,p,\sigma}$, $n_{i,p}=n_{i,p,\uparrow}+ n_{i,p,\downarrow}$, $i$ is the site
index, $p$  the chain index, $t$ the intrachain hopping,
$t_\perp$ the interchain hopping, $U$  the on-site repulsion,
$V_\parallel$ the nearest-neighbor repulsion,  and $V_\perp$  the
interchain  repulsion.

 Since we will be focusing on
 the case of very strongly repulsive interactions, the
single-particle interchain hopping term $t_\perp$ will be
irrelevant in this regime. However, it is well known that even when
$t_\perp$ is irrelevant, it generates an interchain exchange term in
the Hamiltonian\cite{brazovskii_transhop,schulz_2chains}.
 Thus, the Hamiltonian~(\ref{eq:latt-hubbard}) should also
include an interchain exchange term,
\begin{equation}
  \label{eq:interchain-exchange}
  H_{\text{exch.}}=J_\perp \sum_{i} \vec{S}_{i,1}\cdot S_{i,2},
\end{equation}
\noindent where
$S_{i,p}=c^\dagger_{i,p,\alpha}\vec{\sigma}_{\alpha,\beta}c_{i,p,\beta}$.
 In the limit of $t_\perp \ll U$, $J_\perp$ is
estimated to second order perturbation theory in $t_\perp/U$ as
$J_\perp \sim t_\perp^2/U$. In the following, we will consider the
Hamiltonian (\ref{eq:latt-hubbard}) completed by the term
(\ref{eq:interchain-exchange}) and neglect altogether the
interchain hopping $t_\perp$. In fact, another model of interest
is the two leg t-J ladder at quarter filling. The Hamiltonian of this
model reads:
\begin{eqnarray}
  \label{eq:tJ-hamiltonian}
  H&=&-t \sum_{i,p,\sigma} {\mathcal P}(c^\dagger_{i+1,p,\sigma}
  c_{i,p,\sigma}+c^\dagger_{i,p,\sigma} c_{i+1,p,\sigma}){\mathcal P} -t_\perp
  \sum_{i,\sigma} {\mathcal P}(c^\dagger_{i,2,\sigma} c_{i,1,\sigma} +
  \text{H.c.}){\mathcal P} \nonumber \\
  && + \sum_{i,p} (J_\parallel \vec{S}_{i,p}\cdot \vec{S}_{i,p+1} +
  V_\parallel n_{i,p} n_{i+1,p}) + \sum_i  J_\perp \vec{S}_{i,1}\cdot \vec{S}_{i,2} +
  V_\perp \sum_i n_{i,1} n_{i,2},
\end{eqnarray}
\noindent where ${\mathcal P}$ is an operator that projects onto
singly occupied states\cite{tsunegutsu_2ch,white_ladder_friedel}.
We will thus treat $J_\perp$ as a parameter independent of $t,U$  in
(\ref{eq:latt-hubbard})
 in order to be able to discuss also results relevant to the
two-leg $t-J$ ladder model.
In the rest of the present section, we will illustrate two simple
limiting cases of the problem defined by the Hamiltonian
(\ref{eq:latt-hubbard})--~(\ref{eq:interchain-exchange})  which
display insulating
charge ordered ground states.

\subsection{large on-site repulsion}
\label{sec:spinless}

Let us begin by considering the limit $U \to \infty$ in the
Hamiltonian~(\ref{eq:latt-hubbard})--(\ref{eq:interchain-exchange}).
In this limit, it is not possible to put two fermions on the
same site even when
they have opposite spins and from the point of view of
charge excitations, the system behaves as if it was made of spinless
fermions with a Fermi wavevector $k_F'$ twice the one of the spinful fermions
at $U=0$\cite{ogata_inf}. The original spinful fermions system being
quarter-filled, the spinless fermions system is half-filled
and its  effective  Hamiltonian reads:
\begin{eqnarray}
  \label{eq:latt-tV}
  H=-t \sum_{i,p} (a^\dagger_{i+1,p} a_{i,p} + a^\dagger_{i,p}
  a_{i+1,p}) + V_\parallel  \sum_{i,p} n_{i,p} n_{i+1,p} + V_\perp
  \sum_i n_{i,1} n_{i,2},
\end{eqnarray}
\noindent where now $n_{i,p}= a^\dagger_{i,p}a_{i,p}$.
The bosonization of the Hamiltonian (\ref{eq:latt-tV}) is
straightforward. The relevant formulas can be found for instance in
\cite{schulz_houches_revue}.
We obtain the following bosonized Hamiltonian:
\begin{eqnarray}
  \label{eq:bosonized-spinless}
  H=\sum_{r=1,2} \left\{ \int \frac{dx}{2\pi} \left\lbrack uK (\pi \Pi_r)^2 +
  \frac u K  (\partial_x \phi_r)^2\right\rbrack +  + \frac{2g}{(2\pi
a)^2}\int dx \cos 4\phi_r\right\} \frac{V_\perp a}{(\pi a)^2}
\int dx \cos 2\phi_1 \cos 2 \phi_2 + \frac{V_\perp a}{\pi^2}
\int dx \partial_x \phi_1 \partial_x \phi_2,
\end{eqnarray}
\noindent where $\cos 4\phi_{1,2}$ represent the intrachain Umklapp
interactions. Let us first consider the case in which these
processes are irrelevant ($V_\perp=0$). This case corresponds to
$|V_\parallel| <2t$. The Hamiltonian (\ref{eq:latt-tV}) has then a
Luttinger liquid ground state for $V_\perp=0$, and the
renormalized Luttinger exponent $K^*$ can be obtained from studies
of the $t-V$ model \cite{haldane_xxzchain} as:
\begin{eqnarray}
  \label{eq:K-V-t}
  K^*=\frac{1}{2-\frac 2 \pi \arccos \left(\frac {V_\parallel}{2t}\right)}.
\end{eqnarray}
In the Luttinger liquid regime, we can set $g=0$ and replace $K$ by
$K^*$ in the Hamiltonian~(\ref{eq:bosonized-spinless})
provided $V_\perp$ is sufficiently small.
The Hamiltonian~(\ref{eq:bosonized-spinless}), is then decoupled by
introducing the fields $\phi_{\pm}=(\phi_1 \pm \phi_2)/\sqrt{2}$ leading
to the following Hamiltonian:
\begin{eqnarray}
  \label{eq:decoupled-spinless}
&&H=H_++H_-, \nonumber \\
&&  H_r = \int \frac{dx}{2\pi} \left[ u_r K_r (\pi \Pi_r)^2 +
  \frac {u_r} {K_r}  (\partial_x \phi_r)^2\right] + \frac{2V_\perp a} {(2 \pi a)^2} \int dx
 \cos \sqrt{8} \phi_r,
\end{eqnarray}
\noindent where $r=\pm$ and:
\begin{eqnarray}
  \label{eq:constants-spinless}
&& u_r^2=u^2\left(1+ r \frac{KV_\perp a}{\pi u}\right) \nonumber \\
&& K_r^2=K^2 \left(1+ r \frac{KV_\perp a}{\pi u}\right)^{-1}
\end{eqnarray}
For $K_r<1$, the term $\cos \sqrt{8} \phi_r$ is
relevant and opens a gap in the charge modes.
If $V_\perp \ll u/a$, this implies that the
gap opens  as
soon as $K^*<1$, whereas the intrachain Umklapp processes become
relevant only
for $K^*=1/2$. We thus see that for a wide range of
$0<V_\parallel<2t$, although
 intrachain repulsion $V_\parallel$
alone is too weak to open a charge gap by itself,
 the existence of a nonzero interchain repulsion is sufficient to
 induce an insulating ground state. This is consistent with the
 numerical observation  of a charge gap state forming for
 $V_\parallel>0$ and $t_\perp<t$ at large $U$ in
 Ref.~\cite{vojta_qfl}. In that insulating
 state, the charge gap should vary as $\Delta_{\rho+} \sim u/a (V_\perp
 a/u)^{1/(2-K_+)}$.  \\

We note from (\ref{eq:K-V-t}) that $K^*=1$ corresponds
to $V_\parallel=0$, i.e. the Hamiltonian (\ref{eq:decoupled-spinless}) is
identical to the bosonized Hamiltonian of the half-filled 1D Hubbard
chain, $\phi_+$ playing the role of $\phi_\rho$, and $\phi_-$ playing
the role of $\phi_\sigma$. As a result, although the total charge mode
is gapful, the antisymmetric charge mode remains gapless.
 This insulating state presents the same
$SU(2) \times SU(2)\sim SO(4)$ symmetry as the Hubbard model. In
particular, the gapped charge excitations behave like
those of the half-filled Hubbard model and the charge gap is
$\Delta_{\rho+} \sim u/a \exp (-Cu/(V_\perp a))$.

To determine the
long range order that is realized in the ground state, we need to
fix the values of $\langle \phi_{1,2}\rangle$ that minimize the
classical ground state energy, i. e.  require that $\cos
2\langle\phi_1 \rangle \cos 2 \langle \phi_2 \rangle$ is negative.
Since the density of the spinless fermions reads:
\begin{eqnarray}
  \label{eq:spinless-density}
  \rho_p(x)=-\frac 1 \pi \partial_x \phi_p(x) +\frac 1 {\pi a} \cos
  (2\phi_p -2k'_F x),
\end{eqnarray}
\noindent where $p=1,2$ is the chain index, 
this leads to two out of phase charge density
waves of wavevector $2k_F'=\frac \pi a$ on chains 1 and 2, i.e. a
zig-zag charge ordering. This ordering is represented on
figure~\ref{fig:zig-zag}.  We note that in
Ref.~\onlinecite{vojta_qfl}, it was found that the charge ordering was
formed for $U=\infty$ only when $V_\parallel>2t$
although a charge gap was obtained for $V_\parallel>0$.
The reason for this discrepancy in Ref.~\onlinecite{vojta_qfl} could be the
presence of $t/t_\perp\sim 1$.
  Turning to transport properties, from
(\ref{eq:spinless-density})  the long wavelength density on chain
$p$ is $\rho_p(x)=-\frac 1 \pi \partial_x \phi_p$. By considering
the topological
 charge of the sine-Gordon models (\ref{eq:decoupled-spinless}), one
 easily obtains that  a charge solitons carries the electrical charge
 $\pm e$. The a.c. electrical conductivity of this system can be
 obtained from the form factor expansion
 \cite{jeckelmann_hubbard1d,controzzi_mott}.

When $V_\parallel > 2t$, the interchain
Umklapp processes become relevant and induce the formation of charge
density waves on each chain\cite{shankar_spinless_conductivite}. The
effect of $V_\perp$ is to lock the respective phases of these
charge density waves into a zig-zag pattern. Zig-zag charge ordering
is thus ubiquitous in the limit $U=\infty$.

\subsection{large interchain exchange}\label{sec:pairs}
In the limit in which $J_\perp\to \infty$ in
(\ref{eq:latt-hubbard})--~(\ref{eq:interchain-exchange}) or in
(\ref{eq:tJ-hamiltonian}), it is again possible to
derive a simplified Hamiltonian in a low-energy subspace.
Namely, one can restrict to a subspace in which
 fermions form spin singlet pairs on the same rung. These
 pairs can then be treated as hardcore
bosons\cite{troyer_2ch_tj} moving on a single chain and
carrying a charge $2e$.
For a quarter filled fermion system, the effective boson system is at
half-filling. The effective bosonic Hamiltonian reads:
\begin{eqnarray}
  \label{eq:eff-large-Jp}
  H=-\tilde{t} \sum_i (b^\dagger_{i} b_{i+1} + \text{H. c.}) +
  \tilde{V} \sum_i n_i n_{i+1},
\end{eqnarray}
\noindent where $b_i=\sum_\sigma \sigma c_{i,1,\sigma}c_{i,2,-\sigma}$,
annihilates a singlet pair on site $i$. In the limit $J_\perp\gg
t$, the coefficients $\tilde{t},\tilde{V}$ can be obtain
from second order perturbation theory as
$\tilde{t}=t^2/J_\perp$, $\tilde{V}=2t^2/J_\perp+V_\parallel$.
The Hamiltonian (\ref{eq:eff-large-Jp}) is solved exactly by the
Bethe Ansatz\cite{shankar_spinless_conductivite}, and its spectrum has
a gap for $V_\parallel>0$.
A bosonization
description\cite{black_equ,nijs_equivalence,shankar_spinless_conductivite}
of the low energy
excitations of this system gives:
\begin{equation}
  H=\int \frac{dx}{2\pi}\left[ uK (\pi \Pi_b)^2 + \frac u K (\partial_x
    \phi_b)^2\right] -\frac{2\Delta}{(2\pi a)^2} \int dx \cos 4\phi_b,
\end{equation}
where the mode $\phi_b$ describes the charge excitations of the hardcore
boson  system.
The electrical charge density is given by:
\begin{equation}
\rho_e(x)=-\frac {2e}{\pi} \partial_x \phi_b + 2e \frac{e^{i\pi \frac x
    a}}{\pi a} \cos 2\phi_b
\end{equation}
In the gapped regime, $\langle \phi_b \rangle=0$, and
the ground state of this system is formed of singlet pairs in a charge
ordered state as represented on Fig.~\ref{fig:pizerocdw}.
We therefore see that in the regime of strong interaction, both
$J_\perp$ and $V_\perp$ will induce a charge ordering. However,
$J_\perp$ favors a phase with stripes formed along the rungs of the
system, whereas $V_\perp$ favors a zig-zag charge ordering. As a
result, the competition between $J_\perp$ and $V_\perp$ induces a
frustration in the system that can lead to a variety of charge
ordering patterns.

\section{Bosonization description}\label{sec:bosonization}
In the present section, we derive a bosonized representation of the
Hamiltonian (\ref{eq:latt-hubbard})--(\ref{eq:interchain-exchange})
in the limit $J_\perp,V_\perp \ll t$. For
$J_\perp=V_\perp=0$, the chains are decoupled, and
the Hamiltonian describing their low energy, long wavelength excitations
reads:
\begin{eqnarray}
  \label{eq:decoupled-chains}
 H= \sum_{p=1,2} H_{\rho,p}+ H_{\sigma,p} \\
\label{eq:decoupled-charge}
 H_{\rho,p}= \int dx \left[ u_\rho K_\rho (\pi \Pi_{\rho,p})^2 +
    \frac{u_\rho}{K_\rho} (\partial_x \phi_{\rho,p})^2\right] \\
\label{eq:decoupled-spin}
 H_{\sigma,p}= \int dx \left[ u_\sigma K_\sigma (\pi \Pi_{\sigma,p})^2 +
    \frac{u_\sigma}{K_\sigma} (\partial_x \phi_{\sigma,p})^2\right]
\end{eqnarray}

\noindent where the fields satisfy to canonical commutation
relations $[\phi_{\nu,p}(x),\Pi_{\nu',p'}(x')]=i \delta_{\nu,\nu'}
\delta_{p,p'} \delta(x-x')$, ($\nu=\rho,\sigma$).  Since we do not
assume $U,V_\parallel \ll t$, we
 use the renormalized values of $u_\rho,K_\rho,u_\sigma, K_\sigma$
in the bosonized Hamiltonian of the decoupled chains
(\ref{eq:decoupled-charge})-(\ref{eq:decoupled-spin}). Spin rotational invariance
imposes the renormalized value of $K_\sigma=1$. The renormalized values of the
remaining quantities can be obtained from the Bethe
Ansatz\cite{schulz_hubbard_exact,kawakami_tj,kawakami_hubbard,frahm_confinv}
for the Hubbard model or the t-J model at the supersymmetric point, or from numerical
calculations in the case of a non-integrable
model\cite{mila_zotos,sano_extended_hubbard_1d,nakamura_tJ}.

\subsection{derivation of interchain coupling}
\label{sec:finite-U}

In Sec.~\ref{sec:spinless} we have seen that the  $4k_F$ harmonics in the fermion
density play a crucial role for $U/t\gg 1$. The expression of the $4k_F$ harmonics in
terms of the boson fields can be obtained from
Refs.~\onlinecite{haldane_bosons,schulz_wigner_1d}:
\begin{eqnarray}
  \label{eq:density4kF}
  \rho_r(x)=-\frac{\sqrt{2}}{\pi } \partial_x \phi_{\rho_r} +
  \frac{e^{i(\sqrt{2} \phi_{\rho_r} -2 k_F x)}}{\pi a} \cos \sqrt{2}
  \phi_{\sigma_r}(x) + \frac{C}{\pi a}e^{ i 2(\sqrt{2} \phi_{\rho_r} -2 k_F x)},
\end{eqnarray}
\noindent where $r=1,2$ and $k_F=\frac{\pi}{4a}$.
For $U\to \infty$, we recover the bosonized
expression~(\ref{eq:spinless-density})  of the
density of spinless fermions with $k_F'=2k_F$, $\phi=\sqrt{2} \phi_\rho$ and
$K=2 K_\rho$.
Using the expression~(\ref{eq:density4kF}) of the fermion density, we obtain the
bosonized expression of the interchain repulsion $V_\perp$:
\begin{eqnarray}
  \label{eq:interchain}
  H_{\text{V}_\perp}=V_\perp a \int dx  \left[\frac{2}{\pi^2} \partial_x
    \phi_{\rho_1} \partial_x \phi_{\rho_2} + \frac 2 {(\pi a)^2} \cos \sqrt{2}
    (\phi_{\rho_1}-\phi_{\rho_2}) \cos \sqrt{2} \phi_{\sigma_1} \cos
    \sqrt{2} \phi_{\sigma_2} + \frac{C^2}{(\pi a)^2} \cos \sqrt{8}
    \phi_{\rho_1}  \cos \sqrt{8}
    \phi_{\rho_2}\right].
\end{eqnarray}

To obtain the bosonized expression of the spin exchange interaction, we need the spin
density operators:
\begin{eqnarray}
  \label{eq:Sx}
  S^x_p(x)&=&\frac 1 {\pi a} \cos \sqrt{2} \theta_{\sigma,p}(x)\cos \sqrt{2} \phi_{\sigma,p}(x) + \frac
  {e^{i\sqrt{2} \phi_{\rho,p} -2k_F x}} {2\pi a}   \cos \sqrt{2}
  \theta_{\sigma,p}(x), \\
\label{eq:Sy}
  S^y_p(x)&=&\frac 1 {\pi a} \sin \sqrt{2} \theta_{\sigma,p}(x)\cos \sqrt{2} \phi_{\sigma,p}(x) + \frac
  {e^{i\sqrt{2} \phi_{\rho,p} -2k_F x}} {2\pi a}   \sin \sqrt{2}
  \theta_{\sigma,p}(x), \\
\label{eq:Sz} S^z_p(x)&=&- \frac 1 {\pi \sqrt{2}} \partial_x \phi_{\sigma,p} + \frac
  {e^{i\sqrt{2} \phi_{\rho,p} -2k_F x}} {2\pi a}   \sin \sqrt{2}
  \phi_{\sigma,p}(x),
\end{eqnarray}
\noindent where $\theta_{\nu,p}=\pi \int^x \Pi_{\nu,p}(x') dx'$. Since the terms
coming from the $4k_F$ harmonics are less relevant than those produced by the $2k_F$
harmonics, in Eqs.~(\ref{eq:Sx})--(\ref{eq:Sz}) we have altogether neglected the
$4k_F$ harmonics, whose expression can be found in the
Appendix~\ref{app:phen-spin-dens}. From the bosonized expression of the spin
densities, we obtain the exchange interaction as:
\begin{eqnarray}
  \label{eq:exch-int-boson}
  H_{\text{J}_\perp}&& =\frac{J_\perp}{2\pi^2}
  \int dx  \partial_x
  \phi_{\sigma,1}  \partial_x
  \phi_{\sigma,2} + \frac{J_\perp}{2(\pi a)^2} \int dx \cos
  \sqrt{2}(\theta_{\sigma,1}-\theta_{\sigma,2})
  \cos\sqrt{2}(\phi_{\sigma,1}+\phi_{\sigma,2}) \nonumber \\
  && + \frac{J_\perp}{(2\pi
    a)^2} \int dx \cos\sqrt{2}(\phi_{\rho,1}-\phi_{\rho,2}) \left[ 2 \cos
  \sqrt{2}(\theta_{\sigma,1}-\theta_{\sigma,2}) + \cos
  \sqrt{2}(\phi_{\sigma,1}-\phi_{\sigma,2}) -
  \sqrt{2}(\phi_{\sigma,1}+\phi_{\sigma,2})\right] .
\end{eqnarray}

The total interchain interaction Hamiltonian $H_\perp= H_{\text{V}_\perp}+ H_{\text{J}_\perp}$,
is more conveniently written by introducing the new canonically conjugate
fields\cite{khveshenko_2chain} $\phi_{\nu,\pm}=(\phi_{\nu,1}\pm\phi_{\nu,2})/\sqrt{2}$
and $\Pi_{\nu,\pm}=(\Pi_{\nu,1}\pm\Pi_{\nu,2})/\sqrt{2}$, as:
\begin{eqnarray}
  \label{eq:full}
  H_\perp &=&\frac{(4V_\perp+J_\perp)a}{(2\pi a)^2} \int dx
  \cos 2 \phi_{\rho-} \cos 2\phi_{\sigma-} + \frac{(4V_\perp-J_\perp)a}{(2\pi a)^2} \int dx
  \cos 2 \phi_{\rho-} \cos 2\phi_{\sigma+} +  \frac{2J_\perp a}{(2\pi a)^2} \int dx
  \cos 2 \phi_{\rho-} \cos 2\theta_{\sigma-} \nonumber \\
  &+&  \frac{2C^2 V_\perp a}{(2\pi a)^2} \int dx
  (\cos 4 \phi_{\rho-}+\cos 4\phi_{\rho+}) + \frac{V_\perp}{\pi^2}
  \int dx \left[(\partial_x \phi_{\rho+})^2-(\partial_x
    \phi_{\rho-})^2\right] + \frac{J_\perp}{4\pi^2} \int dx \left[(\partial_x \phi_{\sigma+})^2-(\partial_x
    \phi_{\sigma-})^2\right].
\end{eqnarray}

 From the bosonized expression (\ref{eq:full}) of the interchain
interactions and the Hamiltonians of the decoupled chains
(\ref{eq:decoupled-charge})-(\ref{eq:decoupled-spin}) we obtain the
full bosonized form of the Hamiltonian
(\ref{eq:latt-hubbard})-(\ref{eq:interchain-exchange}). 
The total charge mode $\phi_{\rho+}$ decouples from the spin and
staggered charge modes. Below, we will start discussing the properties of the
Hamiltonians of these modes.

\subsection{total charge Hamiltonian}
\label{sec:total-charge-hamilt}
The total charge excitation Hamiltonian is:
\begin{eqnarray}
  \label{eq:charge-hamiltonian}
  H_{\rho+}=\int \frac{dx}{2\pi} \left[ u_{\rho+} K_{\rho+} (\pi
    \Pi_{\rho+})^2 + \frac{u_{\rho+}}{ K_{\rho+}} (\partial_x
    \phi_{\rho+})^2 \right] + \frac{2g_0}{(2\pi a)^2} \int dx \cos 4
  \phi_{\rho+},
\end{eqnarray}
\noindent where:
\begin{eqnarray}
  \label{eq:coeff-rho+}
&&  u_{\rho+}=u_\rho \left(1+\frac{2K_\rho V_\perp a}{\pi
      u_\rho}\right)^{\frac 1 2} \\
&&  K_{\rho+}=K_\rho \left(1+\frac{2K_\rho V_\perp a}{\pi
      u_\rho}\right)^{-\frac 1 2} \\
&&  g_0=C^2 V_\perp
\end{eqnarray}
\noindent The expression of the interchain coupling $g_0$ in
(\ref{eq:charge-hamiltonian}) can also be derived in perturbation
theory by the approach of Ref.~\onlinecite{tsuchiizu_qf1d} and a
sketch of such derivation can be found in the Appendix
\ref{app:deriv-umkl-term}. We note that the same sine-Gordon
Hamiltonian was derived in Sec.~\ref{sec:spinless} in the limit
$U/t\to \infty$, with the identification $\phi_+=\sqrt{2}\phi_\rho$,
$K=2K_\rho$, $K_+=2K_{\rho+}$, and $C=1$. A similar Hamiltonian was
also obtained in Sec.~\ref{sec:pairs} in the limit $J_\perp \to
\infty$. This fact indicates a continuity relation between the weak
and the strong coupling regimes for the charge excitations. Finally,
the Hamiltonian for the total charge mode
(\ref{eq:charge-hamiltonian}) can also be recovered from a general
argument\cite{oshikawa_plateaus,white_ladder_friedel}:
 From Eq.~(\ref{eq:density4kF}), it is easily seen that a translation
by one lattice site amounts to making $\sqrt{2} \phi_{\rho,n}
\to\sqrt{2} \phi_{\rho,n} -\pi/2$, and thus $\phi_{\rho+} \to
\phi_{\rho+}-\pi/2$. Translation invariance of the lattice Hamiltonian
requires the continuum bosonized Hamiltonian to be invariant under
such transformation. The most relevant Umklapp operator compatible
with this symmetry is $\cos 4 \phi_{\rho+}$, and it should be the
operator responsible for the opening of the gap in the charge modes at
quarter-filling.  As a result, in the strong coupling regime, we find
a boundary between the gapped and the gapless regime given by
$K_{\rho+}=1/2$. Concerning the intrachain Umklapp terms, they are of
the form \cite{schulz_mott_revue,giamarchi_mott_shortrev} $\cos 4
\sqrt{2} \phi_{1,2}$ and become relevant only for $K_\rho<1/4$. Thus
we can safely neglect them compared with interchain Umklapp terms.
A final remark is in order concerning the sign of $g_0$. In
appendix~\ref{app:phen-spin-dens}, we show that the interaction in the
spin sector only produces a term of the form $\cos 4\phi_{\rho+} \cos
2\theta_{\sigma-} \cos 2\phi_{\sigma+}$. This latter term is less
relevant than a term of the form $\cos 4\phi_{\rho+}$, so that we can
usually neglect it and assume $g_0>0$.
However, if the spin gap in the system is large,
this term cannot be neglected and contributes a correction  $\cos
4\phi_{\rho+} \langle \cos
2\theta_{\sigma-} \cos 2\phi_{\sigma+}\rangle$ to the Hamiltonian
(\ref{eq:charge-hamiltonian}). This can result in a change of the sign
of $g_0$. In that case, instead of having $\langle \phi_{\rho+}\rangle=\frac
\pi 4$ in the ground state, we should have instead $ \langle
\phi_{\rho+}\rangle=0$. In the following we will discuss both the phases with
$\langle \phi_{\rho+}\rangle=\frac \pi 4$ and $\langle \phi_{\rho+}\rangle=0$. The
latter ones are expected to be obtained when $J_\perp\gg V_\perp$
since we need a large spin gap to modify the sign of $g_0$.

\subsection{spin and charge difference Hamiltonian}

 From  Eqs.~(\ref{eq:decoupled-charge})-(\ref{eq:decoupled-spin})
 and Eq.~(\ref{eq:full}), we obtain the
Hamiltonian describing the interaction of spin modes $\phi_{\sigma \pm}$ and
interchain charge modes $\phi_{\rho-}$.
Under renormalization group (RG) transformation, besides the interactions already
present in the bare Hamiltonians, new interactions can be generated. Thus the
Hamiltonian to consider reads:
\begin{eqnarray}
  \label{eq:exch-ham}
   \tilde{H}&=& H_{\rho-}+H_{\sigma+}+H_{\sigma-}=\int \frac{dx}{2\pi} \sum_{\nu \in \{\rho-,\sigma+,\sigma-\}} \left[
  u_\nu K_\nu (\pi \Pi_\nu)^2+ \frac{u_\nu}{K_\nu} (\partial_x
  \phi_\nu)^2 \right]\nonumber \\
&+& \frac{2g_1}{(2\pi a)^2} \int dx \cos 2\phi_{\sigma+} \cos
2\phi_{\sigma-} + \frac{2g_2}{(2\pi a)^2} \int dx \cos 2
\phi_{\sigma+} \cos 2\theta_{\sigma-} + \frac{2g_3}{(2\pi a)^2} \int dx \cos 4
\phi_{\rho-}\nonumber \\ & +& \frac{2g_4}{(2\pi a)^2}\int dx \cos
2\phi_{\rho-} \cos 2 \phi_{\sigma+} + \frac{2g_5}{(2\pi a)^2}\int dx \cos
2\phi_{\rho-} \cos 2 \phi_{\sigma-} + \frac{2g_6}{(2\pi a)^2}\int dx \cos
2\phi_{\rho-} \cos 2 \theta_{\sigma-},
\end{eqnarray}
\noindent where:
\begin{eqnarray}
  \label{eq:initial-conditions}
&& u_{\rho-}=u_\rho \left(1-\frac {2V_\perp K_\rho a}{\pi
     u_\rho}\right)^{1/2}\\
&& u_{\sigma+}=u_\sigma\left(1+\frac{J_\perp a}{2\pi u_\rho}\right)^{1/2}, \mbox{ }u_{\sigma-}=u_\sigma\left(1-\frac{J_\perp a}{2\pi
    v_F}\right)^{1/2}, \\
&& K_{\rho-}=K_\rho\left(1-\frac {2V_\perp K_\rho a}{\pi
    u_\rho}\right)^{-1/2},\\
&& K_{\sigma+}=\left(1+\frac{J_\perp a}{2\pi v_F}\right)^{-1/2},  \mbox{ } K_{\sigma-}=\left(1-\frac{J_\perp a}{2\pi v_F}\right)^{-1/2}, \\
&&  g_1=0,  \mbox{ } g_2=J_\perp a, \mbox{ } g_3= 2 C^2 V_\perp a,\\
&&  g_4=(2 V_\perp-\frac{J_\perp} 2)a,  \mbox{ } g_5=(2 V_\perp+\frac{J_\perp} 2)a, \mbox{ }g_6=J_\perp a.
\end{eqnarray}

\noindent Note that in Eqs.~(\ref{eq:initial-conditions}), we have put
$g_1=0$ because the intrachain spin interaction in the repulsive
Hubbard model are marginally irrelevant. In full rigor, we should have
added a small marginally irrelevant interaction to the Hamiltonian
(\ref{eq:decoupled-spin}) and we would have obtained $g_1>0$ and
slightly modified expressions of $K_{\sigma\pm}$.
In spite of the fact that we have put $g_1=0$ in the initial
conditions, in the following we will also discuss the case of a
relevant $g_1$ under RG.

To study the Hamiltonian
(\ref{eq:exch-ham}) RG equations can be derived.
Velocity differences between the various modes
can be neglected, and one can take $u_{\sigma\pm}=u_{\rho-}=v_F$.
Introducing $y_i=g_i/(\pi v_F)$ and $K_{\sigma r}=1+y_{\sigma r}/2$, and
using the operator product expansion (OPE) methods\cite{cardy_scaling}, the RG equations read:
\begin{eqnarray}
  \label{eq:full-RGE}
 && \frac{d}{dl}\left(\frac 1 {K_{\rho-}}\right)=y_3^2 + \frac 1 8
  (y_4^2 +y_5^2 +y_6^2), \\
 && \frac{dy_{\sigma-}}{dl}=\frac 1 4 (y_6^2+y_2^2-y_1^2-y_5^2), \\
 && \frac{dy_{\sigma+}}{dl}=-\frac 1 4 (y_1^2+y_2^2+y_4^2), \\
 && \frac{dy_1}{dl}=-\frac 1 2
  \left[(y_{\sigma+}+y_{\sigma-})y_1+y_4y_5\right], \\
  && \frac{dy_2}{dl}=-\frac 1 2
  \left[(y_{\sigma+}-y_{\sigma-})y_2+y_4y_6\right], \\
 && \frac{dy_3}{dl}=(2-4K_{\rho-}) y_3 -\frac 1 8 (y_4^2+y_5^2+y_6^2), \\
 && \frac{dy_4}{dl}=(1-K_{\rho-}) y_4-\frac 1 2 (y_{\sigma+} y_4+y_1
  y_5+ y_2 y_6+y_3 y_4), \\
  && \frac{dy_5}{dl}=(1-K_{\rho-}) y_5-\frac 1 2 (y_{\sigma-} y_5+y_3
  y_5+ y_1 y_4), \\
&& \frac{dy_6}{dl}=(1-K_{\rho-}) y_6+\frac 1 2 (y_{\sigma-} y_6-y_3
  y_6- y_2 y_4).
\end{eqnarray}
\noindent We see from these equations that although $y_1$ is initially
zero, it becomes non-zero under the RG flow. The fact that the problem
under consideration has $SU(2)$ spin rotational invariance leads to a
simplification of the RG equations (\ref{eq:full-RGE}).
The initial conditions (\ref{eq:initial-conditions}) lead to the following
relations  $\forall l$:
\begin{eqnarray}
  \label{eq:hypersurface}
 && y_{\sigma+}(l)+y_{\sigma-}(l)=y_1(l), \\
 && y_{\sigma-}(l)-y_{\sigma+}(l)=y_2(l), \\
 && y_5(l)-y_4(l)=y_6(l).
\end{eqnarray}
These conditions ensure the $SU(2)$ symmetry of the RG flow and reduce
the flow from a curve in a nine-dimensional space to a
curve in a six-dimensional hyperplane. The simplified RG equations now reads:

\begin{eqnarray}
&& \frac{d}{dl}\left(\frac 1 {K_{\rho-}}\right)=y_3^2 + \frac 1 4
  (y_4^2 +y_6^2 +y_4y_6) \\
&&  \frac{dy_1}{dl}=-\frac 1 2
  (y_1^2+y_4^2+y_4y_6) \\
  && \frac{dy_2}{dl}=\frac 1 2
  (y_2^2-y_4y_6) \\
 && \frac{dy_3}{dl}=(2-4K_{\rho-}) y_3 -\frac 1 4 (y_4^2+y_6^2+y_4y_6) \\
 && \frac{dy_4}{dl}=(1-K_{\rho-}) y_4-\frac 1 4 (
 3y_1 y_4-y_2 y_4+ 2y_2 y_6+2y_1y_6+2y_3 y_4) \\
 \label{rgred}
&& \frac{dy_6}{dl}=(1-K_{\rho-}) y_6+\frac 1 4 (y_1y_6+y_2y_6-2y_3y_6-2y_2 y_4).
\end{eqnarray}
The flow of these equations will be discussed in Sec.~\ref{sec:numerics}. Here, we
want to discuss briefly the possible fixed points within a semiclassical argument,
i.e. by considering the expectation values of the phase fields that minimize the
classical ground state energy. By looking at the possible classical minima, we can
distinguish two regimes.
 In the first regime, $\langle
\phi_{\rho-}\rangle=0,\frac\pi 2$ and $\langle \cos 2 \phi_{\rho-}\rangle \ne 0$. This
case is similar to that obtained in ladders at incommensurate
filling\cite{khveshenko_2chain,schulz_2chains,balents_2ch}. In this regime the
relevance of the terms $g_{4,5,6}$ is responsible for the presence of the spin gap. In
the second regime,  $\langle \phi_{\rho-}\rangle=\pm \frac \pi 4$, so that
$\langle\cos 4\phi_{\rho-}\rangle \ne 0$, and $\langle \cos 2 \phi_{\rho-}\rangle=0$.
This regime corresponds to $g_3$ being the dominant interaction, and $g_{4,5,6}$ being
less relevant interactions. This case is specific of the quarter-filled ladder, and
corresponds in the limit $U=\infty$ of Sec.~\ref{sec:spinless} to the formation of the
gap in $\phi_-$. We note that in the present regime, spin gaps can also be generated
by the $g_{1,2}$ terms. When $g_1$ is the relevant interaction, the corresponding
bosonized expression can be rewritten as $\sim g_1 (\cos \sqrt{8} \phi_1 + \cos
\sqrt{8} \phi_2)$, so that the spin gaps correspond to two independent intrachain spin
gaps. We also note that in order to preserve spin
 rotational invariance, we need to have $g_1<0$ in that case.

It is possible to give a roughly estimate of the parameters region where each regime
dominate by looking at the scaling dimensions of the operators. The operator $\cos
4\phi_{\rho-}$ has scaling dimension $4K_{\rho-}$ so that it is relevant for
$K_{\rho-}<1/2$. The operators $\cos 2\phi_{\rho-}\cos 2\phi_{\sigma-}$, $\cos
2\phi_{\rho-}\cos 2\phi_{\sigma+}$, $\cos 2\phi_{\rho-}\cos 2\theta_{\sigma-}$ have
respective scaling dimensions $K_{\rho-}+K_{\sigma-}$, $K_{\rho-}+K_{\sigma+}$,
$K_{\rho-}+1/K_{\sigma-}$. Taking into account the spin rotational symmetry, this
implies that these operators have all the dimension $K_{\rho-}+1$ and become relevant
for $K_{\rho-}<1$. The operator $\cos 4\phi_{\rho-}$ becomes the most relevant
operator when $4K_{\rho-}<1+K_{\rho-}$, i. e. for $K_{\rho-}<1/3$. Therefore, we
expect to have the first regime in the limit of a repulsion not too strong,
$1/3<K_{\rho-}<1/2$, and the second regime in the case of a stronger repulsion
$1/4<K_{\rho-}<1/2$.

\section{Phase diagram}
\label{sec:phase}

In this section, we describe the various insulating phases predicted from the
renormalization group equations (\ref{rgred}). In order to
describe the possible ground states in the phase diagram, we introduce first the corresponding
order parameters.

\subsection{order parameters}

Since for strong repulsion the system has
a gap in the spin excitations, the possible order parameters can only be bond-order
waves (BOW) and charge density waves (CDW). We will denote the corresponding phases as
$(q_x,q_y)-$BOW, and $(q_x,q_y)-$CDW where $q_x=\pi/2$ or $\pi$ and $q_y=0,\pi$ or
$q_y=\pm \pi/2$ if $q_x=\pi/2$. The phases with
a $q_x=\pi$ ordering have at least two-fold degenerate ground state, while
the phases with a  $q_x=\frac\pi 2$  have a four-fold degenerate ground
state.

On the lattice, these operators are defined as:
\begin{eqnarray}
  \label{eq:cdw-fermion-def}
&&  O_{CDW(\pi,q_y)}(i)= (-)^i (n_{i,1}+\cos(q_y) n_{i,2})
(q_y=0,\pi) \\ &&  O_{BOW(\pi,q_y)}(i)= (-)^i \sum_\sigma
(c^\dagger_{i+1,1,\sigma}c^\dagger_{i+1,1,\sigma}+\cos(q_y)c^\dagger_{i+1,2,\sigma}c^\dagger_{i+1,2,\sigma})
(q_y=0,\pi) \\
&&  O_{CDW(\frac \pi 2, q_y)}(i)= e^{-i\frac \pi 2 i} (e^{i q_y/2}
  n_{i,1}+ e^{-i q_y/2} n_{i,2}) (q_y=0,\frac \pi 2, \pi, - \frac \pi
  2).
\end{eqnarray}
The corresponding bosonized expressions is obtained by those of the charge densities
as a function of $\phi_{\rho\pm}$ and $\phi_{\sigma 1,2}$:

\begin{eqnarray}
\label{eq:dens1}
&& \rho_1(x)= -\frac 1 \pi \partial_x (\phi_{\rho_+}+\phi_{\rho_-})
+\frac{e^{i(\phi_{\rho_+}+\phi_{\rho_-}-2k_F x)}}{\pi a} \cos \sqrt{2}
 \phi_{\sigma_1}(x) + \frac{C}{\pi a}e^{ 2i(\phi_{\rho_+}+\phi_{\rho_-}-2k_F
 x)}, \\
 \label{eq:dens2}
&& \rho_2(x)= -\frac 1 \pi \partial_x (\phi_{\rho_+}-\phi_{\rho_-})
+\frac{e^{i(\phi_{\rho_+}-\phi_{\rho_-}-2k_F x)}}{\pi a} \cos \sqrt{2}
 \phi_{\sigma_2}(x) + \frac{C}{\pi a}e^{ 2i(\phi_{\rho_+}-\phi_{\rho_-}-2k_F
 x)}.
\end{eqnarray}

\noindent We consider first $q_x=\pi$. The charge density wave order parameters are
straightforwardly obtained from (\ref{eq:dens1})-(\ref{eq:dens2}) as:
\begin{eqnarray}
  \label{eq:cdw-pi-zero}
  O_{CDW(\pi,0)}&\sim&\frac1 {2\pi\alpha} \cos 2\phi_{\rho+}\cos
  2\phi_{\rho-}\\
\label{eq:cdw-pi-pi}
  O_{CDW(\pi,\pi)}&\sim&\frac1 {2\pi\alpha} \sin 2\phi_{\rho+}\sin
  2\phi_{\rho-}
\end{eqnarray}
The bond order wave order parameters measure the charge density between the sites $i$
and $i+1$, i.e. on site $i+1/2$, and their bosonized expression reads:
\begin{eqnarray}
  \label{eq:bow-pi-zero}
  O_{BOW(\pi,0)}&\sim&\frac1 {2\pi\alpha} \sin 2\phi_{\rho+}\cos
  2\phi_{\rho-}\\
\label{eq:bow-pi-pi}
  O_{BOW(\pi,\pi)}&\sim&\frac1 {2\pi\alpha} \cos 2\phi_{\rho+}\sin
  2\phi_{\rho-}
\end{eqnarray}
Concerning the $\pi/2$ charge density wave order parameter, we have to consider
whether $\phi_{\sigma-}$ or $\theta_{\sigma-}$ is ordered. In the case
$\theta_{\sigma-}$ is ordered, the operators $\cos \sqrt{2}\phi_{\sigma,1,2}$ have
zero expectation value, and exponentially decaying correlations, so that all of the
$(\pi/2,q_y)$-CDW order parameters have short-range order. When $\phi_{\sigma-}$ and
$\phi_{\sigma+}$ are ordered, the operators $\cos \sqrt{2}\phi_{\sigma,1,2}$ have
non-zero expectation values. In this case, the expression of the order parameters for
$(\pi/2,q_y)-$CDW as a function of $\phi_{\rho+}$ and $\phi_{\rho-}$ is:

\begin{eqnarray}
\label{eq:cdw-pi2-zero}
 O_{CDW(\frac \pi 2,0)}& \sim & \frac1 {2\pi\alpha}e^{i\phi_{\rho+}} \cos \phi_{\rho-} \\
\label{eq:cdw-pi2-pi2}
 O_{CDW(\frac \pi 2,\frac \pi 2)} & \sim & \frac1 {2\pi\alpha} e^{i\phi_{\rho+}} \cos(
 \phi_{\rho-}-\frac \pi 4) \\
\label{eq:cdw-pi2-pi}
 O_{CDW(\frac \pi 2,\pi)}& \sim & \frac1 {2\pi\alpha}e^{i\phi_{\rho+}} \sin \phi_{\rho-}
 \\
\label{eq:cdw-pi2-mpi2}
 O_{CDW(\frac \pi 2, -\frac \pi 2)} & \sim & \frac1 {2\pi\alpha} e^{i\phi_{\rho+}} \cos(
 \phi_{\rho-}+\frac \pi 4)
\end{eqnarray}

 From the knowledge of the order parameters, in the following
we are going to discuss the ground states that are realized
for the various possible ordering of the phase fields. We will first
discuss the orderings with $\langle \phi_{\rho-}\rangle=\frac \pi 4$
and then turn to orderings with $\langle \phi_{\rho-}\rangle=0$. As we
have explained in Sec.~\ref{sec:total-charge-hamilt},
the latter type of ordering
should be expected for dominant $J_\perp$.
In Tables \ref{tab:phases-zero}-\ref{tab:phases-pi4}
we give a summary of the phases
corresponding to $\langle \phi_{\rho+}\rangle=\pi/4$ and $\langle
\phi_{\rho+}\rangle=0$. These different phases are detailed in the
forthcoming sections.

\begin{table}
\begin{tabular}{|c|c|c|c|c|}
\hline
  ${\mathbf{\langle \phi_{\rho+} \rangle=0}}$ & $\langle \phi_{\rho-} \rangle$ & $\langle \phi_{\sigma+} \rangle$ & $\langle \phi_{\sigma-} \rangle$ & $\langle \theta_{\sigma-} \rangle$ \\
\cline{1-2}
\cline{2-3}
\cline{3-4}
\cline{4-5}
 \begin{tabular}{c}
 $(\frac{\pi}{2},\pi)$-CDW \\
  $(\pi,0)$-CDW
 \end{tabular}  & $\pm \frac{\pi}{2}$ & 0 & 0 & -- \\
\hline
  $(\pi,0)$-CDW  & $\pm \frac{\pi}{2}$ & 0 & -- & 0 \\
\hline
 $(\pi,\pi)$-BOW  & $\pm \frac{\pi}{4}$ & 0 & -- & 0 \\
\hline
 \begin{tabular}{c}
 $(\frac{\pi}{2},\frac{\pi}{2})$-CDW \\
  $(\pi,\pi)$-BOW
 \end{tabular}  & $\pm \frac{\pi}{4}$  &  $\frac{\pi}{2}$ & 0 & -- \\
\hline
\end{tabular}
\caption{The phases of the quarter-filled ladder in the case $\langle
  \phi_{\rho+} \rangle=0$}
\label{tab:phases-zero}
\end{table}

\begin{table}
\begin{tabular}{|c|c|c|c|c|}
\hline
  $\langle \phi_{\rho+} \rangle=\frac{\pi}{4}$  & $\langle \phi_{\rho-} \rangle$ & $\langle \phi_{\sigma+} \rangle$ & $\langle \phi_{\sigma-} \rangle$ & $\langle \theta_{\sigma-} \rangle$ \\
\hline
\begin{tabular}{c}
 $(\frac{\pi}{2},\pi)$-CDW \\
  $(\pi,0)$-BOW
\end{tabular} & $\pm\frac{\pi}{2}$ & 0 & 0 & -- \\
\hline
 $(\pi,0)$-BOW & $\pm\frac{\pi}{2}$ & 0 & -- & 0 \\
\hline
$(\pi,\pi)$-CDW  & $\pm \frac{\pi}{4}$ & 0 & -- & 0 \\
\hline
\begin{tabular}{c}
$(\frac \pi 2, \frac \pi 2)$-CDW \\
$(\pi,\pi)$-CDW
\end{tabular} & $\pm \frac{\pi}{4}$ & 0 & 0 & -- \\
\hline
\end{tabular}
\caption{The phases of the quarter-filled ladder in the case $\langle
  \phi_{\rho+} \rangle=\frac \pi 4$}
\label{tab:phases-pi4}
\end{table}

\subsection{$(\pi,\pi)$-charge density wave}
\label{sec:zig-zag}

\begin{figure}
\includegraphics[width=\figwidth]{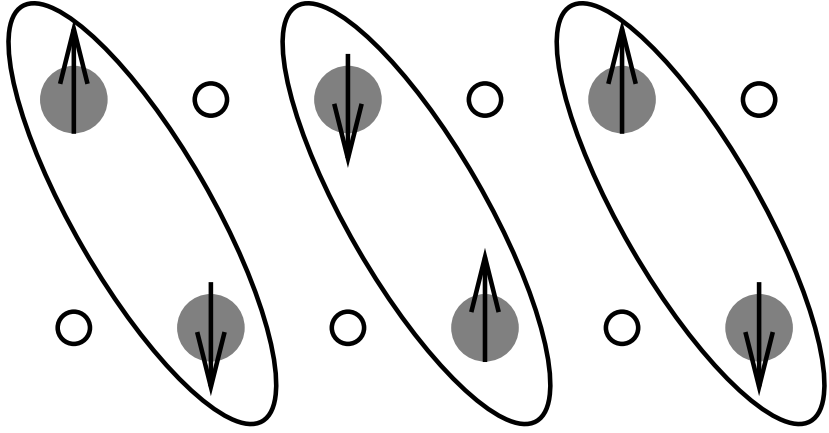}
\caption{\label{fig:zig-zag} The phase with zig-zag charge ordering
and spin gap- $(\pi,\pi)$-CDW. The grey circles represent the local charge density and the arrows
the spins. The empty circles indicate the absence of spin or charge on that site.}
\end{figure}

In the case of $\langle\phi_{\rho+}\rangle=\frac \pi 4$,
$\langle\phi_{\rho-}\rangle=\frac \pi 4$, $\langle \phi_{\sigma+}\rangle=0$, $\langle
\theta_{\sigma-}\rangle=\frac \pi 2$, the $(\pi/2,q_y)-$CDW order parameters have all
zero expectation value and exponentially decaying correlations. Moreover, the
 $(\pi,\pi)$-CDW order parameter has a non-zero expectation
value, leading to a zig-zag charge ordering\cite{mostovoy_nav2o5,vojta_qfl}. This ordering is
represented on Fig.~\ref{fig:zig-zag}.
This charge ordered phase
possesses also a spin gap caused by interchain coupling and identifies
with the CDW$_{\mathrm{sg}}$ (with spin gap) of
Ref.~\onlinecite{vojta_qfl}. The ground state is this phase
is fourfold degenerate (twofold due to translation symmetry, and
twofold due to the two possible orientations of the spin singlets).
We note that this phase has previously
bee discussed in Sec.~\ref{sec:spinless} in the limit of $U\to
\infty$, however in that limit it was not possible to discuss the spin
modes. From the discussion of Sec.~\ref{sec:spinless}, we see that the
essential ingredient for the zig-zag ordering is the mutual locking of the
$4k_F$ charge density fluctuations so that zig-zag ordering needs both
strong intrachain and interchain repulsion. The formation of the spin gap
appears to be unrelated to the zig-zag ordering, but only a
consequence of the coupling of the $q=0$ fluctuations. We note however
that the higher order terms derived in App.~\ref{app:phen-spin-dens}
could lead in the case of a large charge gap to corrections to the
spin Hamiltonian that could enhance the spin gap.
In the RG study of Eqs.~(\ref{eq:full-RGE}), this phase is obtained for $y_3 \to
+\infty$, and $y_2 \to +\infty$.

\subsection{$(\frac\pi 2,\frac\pi 2)$ Charge density wave}

\begin{figure}[htbp]
  \begin{center}
    \includegraphics[width=\figwidth]{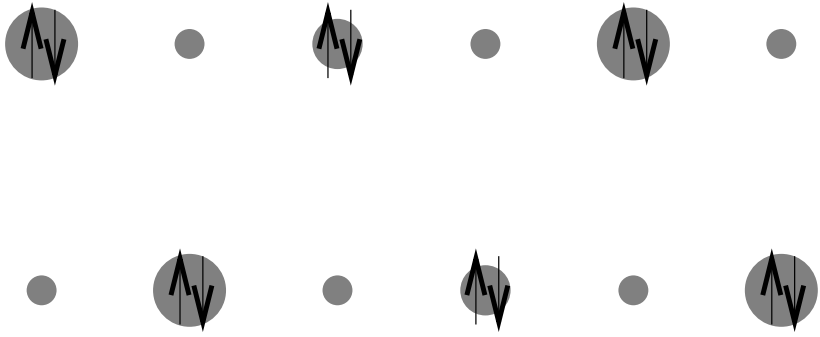}
    \caption{The $(\frac \pi 2,\frac \pi 2)$-charge density wave phase.
The grey circles represent the local charge density. The difference in size
indicates small or large charge density on that site.}
    \label{fig:cdw}
  \end{center}
\end{figure}

In the case of $\langle \phi_{\rho+}\rangle= \langle \phi_{\rho-}\rangle=\pm \frac \pi
4$ and $\langle \phi_{\sigma+}\rangle=0, \langle\phi_{\sigma-}\rangle=0$ a $(\frac \pi
2,\pm \frac\pi 2)$-CDW  state is formed coexisting with a $(\pi,\pi)$-CDW oscillation.
This phase possesses also an intrachain spin gap. The corresponding state is
represented on Fig.~\ref{fig:cdw}. A simple physical picture of this state
 is that a $\frac \pi 2$-charge density is formed in each chain, and
 the interchain repulsion locks the phases of both charge density
 waves. Since there is an intrachain gap, this phase appears to be
 more likely to be observed in a Hubbard ladder with $U<0$ and
 $V_\parallel>0$ or in a t-J ladder in a regime in which $J_\parallel$
 is large enough to cause the formation of a spin gap in the single
 chain\cite{nakamura_tJ}.
 The dephasing between the two charge density waves is $2\langle
\phi_{\rho-}\rangle=\pm \frac \pi 2$. This dephasing results from the
mutual locking of the $4k_F$ density fluctuations, thus pointing to
strong repulsion in the chain. The corresponding phase has a fourfold
degenerate ground state. In the renormalization group treatment, this
phase is obtained for $y_1\to +\infty$ and $y_3 \to +\infty$.

\subsection{$(\pi,0)$ Bond order wave}

\begin{figure}[htbp]
  \begin{center}
    \includegraphics[width=\figwidth]{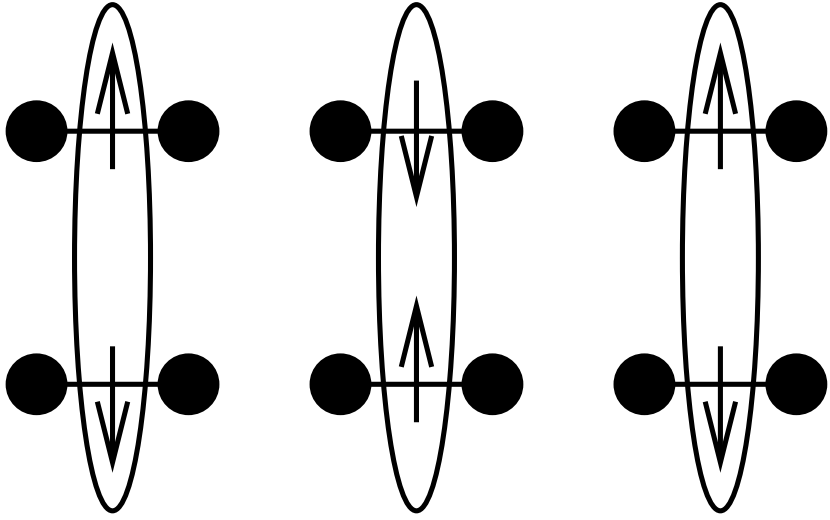}
    \caption{The $(\pi,0)$-bond-order wave phase. The connected black
      circles represent a bond occupied by an electron.}
    \label{fig:bondorder}
  \end{center}
\end{figure}

For $\langle \phi_{\rho+}\rangle=\frac \pi 4$, $\langle
\phi_{\rho-}\rangle=\frac \pi 2$, $\langle \phi_{\sigma+}\rangle=0$,
$\langle \theta_{\sigma-}\rangle=0$, all of the CDW order parameters
vanish. The only nonvanishing order parameter is $(\pi,0)$-BOW. This
BOW can be described in physical terms in the following way: the
fermions are localized on the bonds between two sites, and due to the
$J_\perp$ interaction, they form a spin singlet with the fermion on
the opposite chain. Clearly, this phase should be expected at strong
$J_\perp$ and moderate repulsion in the chains. The ground state is
only twofold degenerate. In the
renormalization group, this phase is obtained for $y_4\to -\infty$,
$y_5+y_6 \to +\infty$, and the inspection of the expression of $y_4$
confirms that $J_\perp$ is the dominant interaction in the
$(\pi,0)$-BOW.
 The corresponding phase is drawn on Figure~\ref{fig:bondorder}.
We note that since this phase has a uniform charge density, it is a
possible candidate for the homogeneous insulator HI$_{\mathrm{sg}}$ phase of
Ref.~\onlinecite{vojta_qfl}. It is also interesting to remark that
this phase is a two-leg analog of the SDW$_1$ phase obtained in
\cite{riera_coexistence_1d} in a system of coupled quarter filled
extended Hubbard chains.

\subsection{$(\frac\pi 2,\pi)$ Charge density wave}
\begin{figure}[htbp]
  \begin{center}
    \includegraphics[width=\figwidth]{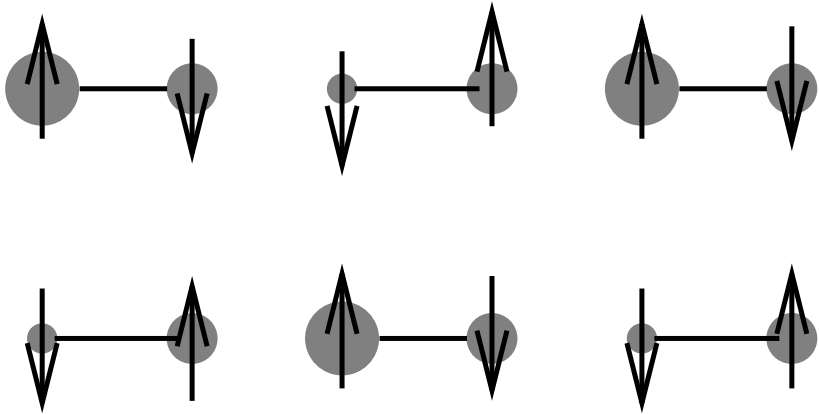}
    \caption{The $(\frac \pi 2,\pi)$-CDW with $(\pi,0)$-BOW. The grey
      circles represent the charge density in Fig.\ref{fig:cdw} and
      the black horizontal lines denote occupied bonds as in
      Fig.~\ref{fig:bondorder}. In contrast with the pure
      $(\pi,0)$-BOW the charge density is inhomogeneous along the chains.}
    \label{fig:pi2picdw1}
  \end{center}
\end{figure}

For $\langle \phi_{\rho+}\rangle=\frac \pi 4$, $\langle \phi_{\rho-}\rangle=\frac \pi
2$, $\langle \phi_{\sigma+}\rangle=0$, $\langle \phi_{\sigma-}\rangle=0$, the $(\frac
\pi 2,\pi)$-CDW order parameter does not vanish. The $(\pi,0)$-BOW correlations are
also present in this phase. A sketch of this phase is given in
Fig.~\ref{fig:pi2picdw1}. Similarly to the $(\frac \pi 2,\frac \pi 2)$-CDW
this phase results from the mutual locking of the $2k_F$ charge
density wave fluctuations of the two coupled chains. However, in
contrast to the $(\frac \pi 2,\frac \pi 2)$-CDW this locking is
produced by the coupling of the $2k_F$ density fluctuations. This implies that
the spin gap is formed
as a result of interchain coupling.
In the RG study this phase is obtained for $y_4 \to +\infty$, $y_5+y_6 \to +\infty$.
Using the expression of $y_4$ we see that this phase is dominated by interchain
repulsion in contrast with the $(\pi,0)$-BOW.

\subsection{$(\pi,\pi)$ Bond order wave}
\begin{figure}[htbp]
  \begin{center}
    \includegraphics[width=\figwidth]{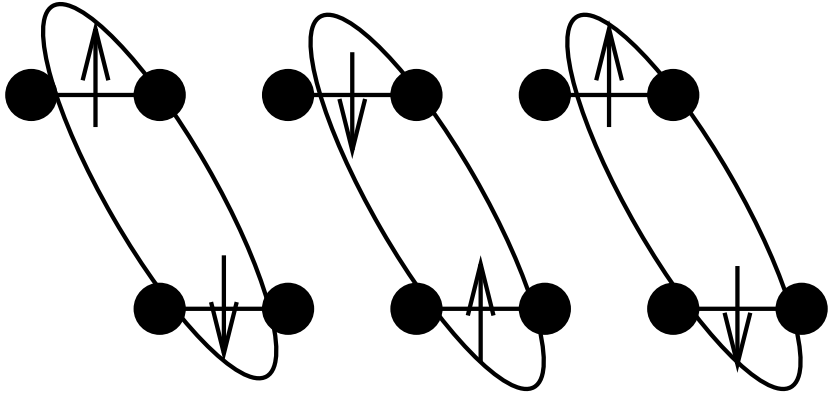}
    \caption{Sketch of the $(\pi,\pi)$-BOW.}
    \label{fig:pipibow}
  \end{center}
\end{figure}

When $\langle \phi_{\rho+}\rangle =0$,$\langle \phi_{\rho-}\rangle =\pm \frac \pi 4$,
$\langle\phi_{\sigma+}\rangle=\frac \pi 2$, $\langle\theta_{\sigma-}\rangle=0$, all of
the CDW order parameters vanish and only  the $(\pi,\pi)$-BOW order parameter is
non-zero. This phase corresponds to a staggered bond ordering shown on
Fig.~\ref{fig:pipibow}. It can be viewed as a $(\pi,\pi)$-CDW
translated by half a lattice spacing. Elementary excitations in this
phase are thus similar to those of the $(\pi,\pi)$-CDW
phase.
 In the renormalization group such state is
obtained for $y_2 \to +\infty$, $y_3 \to +\infty$.

\subsection{$(\frac \pi 2,\frac \pi 2)$ Charge Density Wave}
\begin{figure}[htbp]
  \begin{center}
    \includegraphics[width=\figwidth]{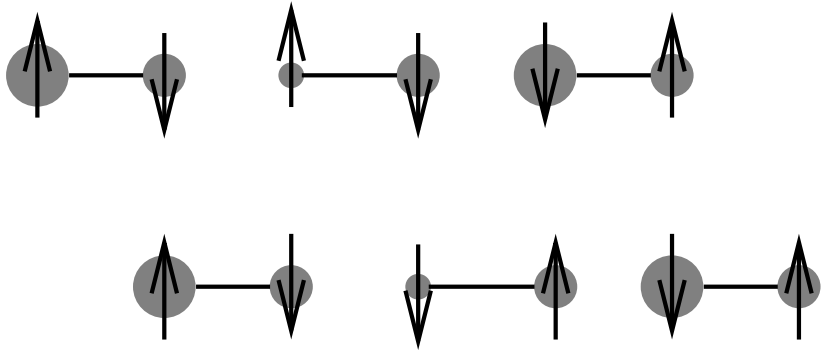}
    \caption{The $(\frac \pi 2,\frac \pi 2)$-CDW with $(\pi,\pi)$-BOW. The grey circles have the same meaning as in Fig.\ref{fig:cdw}. }
    \label{fig:pi2pi2cdwpipibow}
  \end{center}
\end{figure}

 When $\langle \phi_{\rho+}\rangle =0$,$\langle \phi_{\rho-}\rangle
=\pm \frac \pi 4$, $\langle\phi_{\sigma+}\rangle=\frac \pi 2$,
$\langle\phi_{\sigma-}\rangle=0$, the $(\frac \pi 2,\frac \pi 2)$-CDW order parameter
does not vanish and coexists with a $(\pi,\pi)$-BOW. This phase is different from the
$(\frac \pi 2,\frac \pi 2)$-CDW previously encountered as the previous phase contained
a $(\pi,\pi)$-CDW. There are however similarities between these two
phases since both result from the mutual locking of $4k_F$ components
of density fluctuations combined with an intrachain spin gap. Since
the sign change in $g_0$ results from the formation of an interchain
spin gap, we should expect this phase to have only a rather narrow
domain of existence.
This phase is sketched on Fig.~\ref{fig:pi2pi2cdwpipibow}. In the
renormalization group treatment, it is obtained
when $y_1 \to +\infty$, $y_3 \to +\infty$.

\subsection{$(\pi,0)$ Charge Density Wave}
\begin{figure}[htbp]
  \begin{center}
    \includegraphics[width=\figwidth]{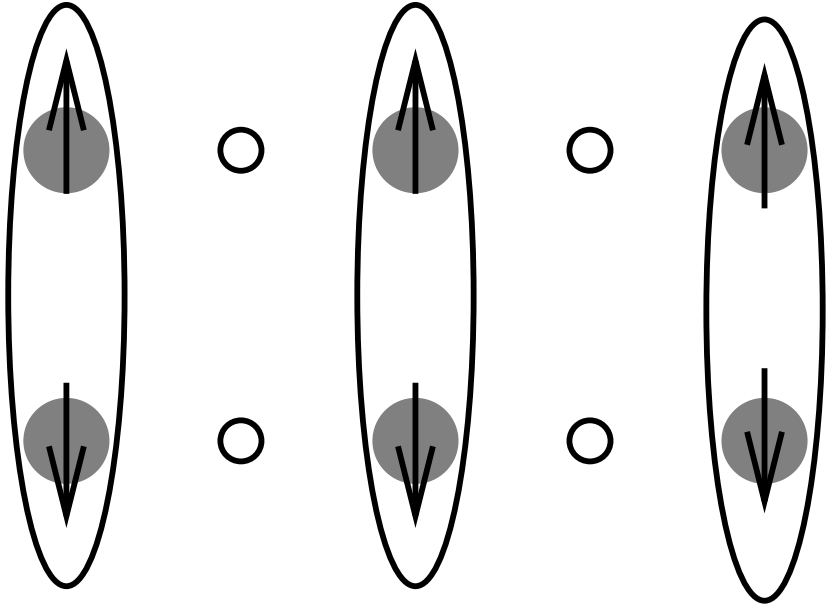}
    \caption{The $(\pi,0)$-CDW. The grey (empty) circles have the same
    meaning as in Fig.\ref{fig:zig-zag}.}
    \label{fig:pizerocdw}
  \end{center}
\end{figure}

When $\langle \phi_{\rho+}\rangle =0$,$\langle \phi_{\rho-}\rangle
=\pm \frac \pi 2$, $\langle\phi_{\sigma+}\rangle=\frac \pi 2$,
$\langle\theta_{\sigma-}\rangle=0$, the only non-vanishing order
parameter is the $(\pi,0)$-CDW one. In the RG study this phase is
obtained for $y_4 \to -\infty$, $y_5+y_6\to +\infty$, i.e. it
corresponds to a dominant $J_\perp$, in agreement with the results of
Sec.~\ref{sec:pairs}. This phase can be viewed as the $(\pi,0)$-BOW
shifted by a half lattice spacing. The corresponding phase is sketched in
Fig.~\ref{fig:pizerocdw}.

\subsection{$(\frac \pi 2,\pi)$ Charge Density Wave}
\begin{figure}[htbp]
  \begin{center}
    \includegraphics[width=\figwidth]{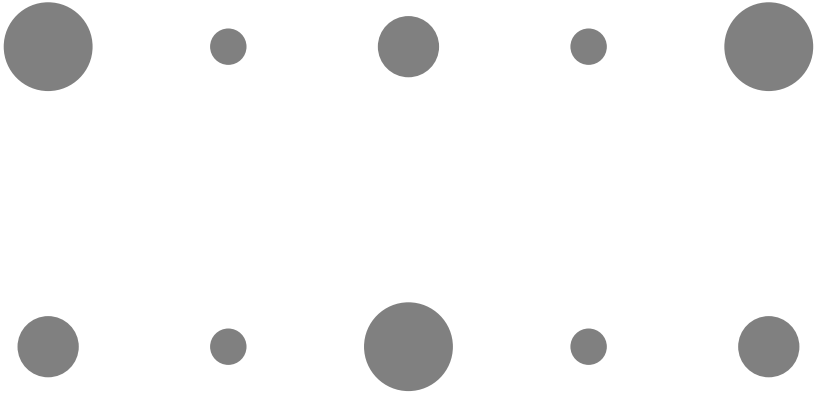}
    \caption{The $(\frac \pi 2,\pi)$-CDW with $(\pi,0)$-CDW.
    The grey circles have the same meaning as in Fig.\ref{fig:cdw}.}
    \label{fig:pi2pi2pizerocdw}
  \end{center}
\end{figure}

When $\langle \phi_{\rho+}\rangle =0$,$\langle \phi_{\rho-}\rangle
=\pm \frac \pi 2$, $\langle\phi_{\sigma+}\rangle=\frac \pi 2$,
$\langle\phi_{\sigma-}\rangle=0$, the only non-vanishing order
parameters are the $(\pi,0)$-CDW and the $(\frac \pi 2,\pi)$-CDW
ones. Under RG study, this phase is obtained for $y_4\to +\infty$,
$y_5+y_6 \to +\infty$ and corresponds to dominant $V_\perp$. For this
reason, we should expect this phase to have a rather narrow domain of
existence. The spin
gap corresponds again to an intrachain spin gap. This phase is
sketched in Fig.~\ref{fig:pi2pi2pizerocdw}.

\section{Elementary excitations}
\label{sec:excitations}
In the present section, we discuss the nature of the elementary
excitations in the various gapped phases of the quarter-filled two leg
ladder. Due to the spin-charge separation, all of the insulating
phases have charge excitations of charge $\pm e$ and zero
spin. However, the nature of the magnetic excitations is dependent of
the nature of the phases considered. Below, we review the various
 phases with their elementary excitations.
\subsection{$(\pi,\pi)$-CDW and $(\pi,0)$-BOW}
\subsubsection{Charge excitations}
In these two phases,  $y_3 \rightarrow +\infty$, implying that the effective
Hamiltonian for the charge sector is described by two sine-Gordon Hamiltonians:

\begin{equation}
H^c_{\text{eff}}=H_+^c+H_-^c,
\end{equation}

\noindent where

\begin{equation}
\label{eq:effham}
H_\alpha^c=\int \frac{dx}{2\pi} \left[ u_{\rho_\alpha} K_{\rho_\alpha} (\pi
   \Pi_{\rho_\alpha})^2 + \frac{u_{\rho_\alpha}}{ K_{\rho_\alpha}} (\partial_x
   \phi_{\rho_\alpha})^2 \right] + \frac{2g_\alpha}{(2\pi a)^2} \int dx \cos 4
 \phi_{\rho_\alpha},
\end{equation}

\noindent $\alpha=\pm$. Thus, the elementary excitations can be
described in terms of solitons of two decoupled sine-Gordon
models. Following a semiclassical argument, a soliton is joining
two consecutive minima of the Hamiltonian (\ref{eq:effham}).
These minimums are given by $<\phi_{\rho_\alpha}>\equiv \pi/4
[\pi/2]$. As a result, the charge of solitons is given by:

\begin{equation}
q_\alpha=-\frac{2}{ \pi} \int_{\infty}^{+\infty} \partial_x \phi_{\rho_\alpha}=
-\frac{2}{ \pi}  \lbrack \phi_{\rho_\alpha}(+\infty)-
\phi_{\rho_\alpha}(-\infty) \rbrack =\pm 1
\end{equation}

\begin{figure}[htbp]
  \begin{center}
    \includegraphics[width=\figwidth]{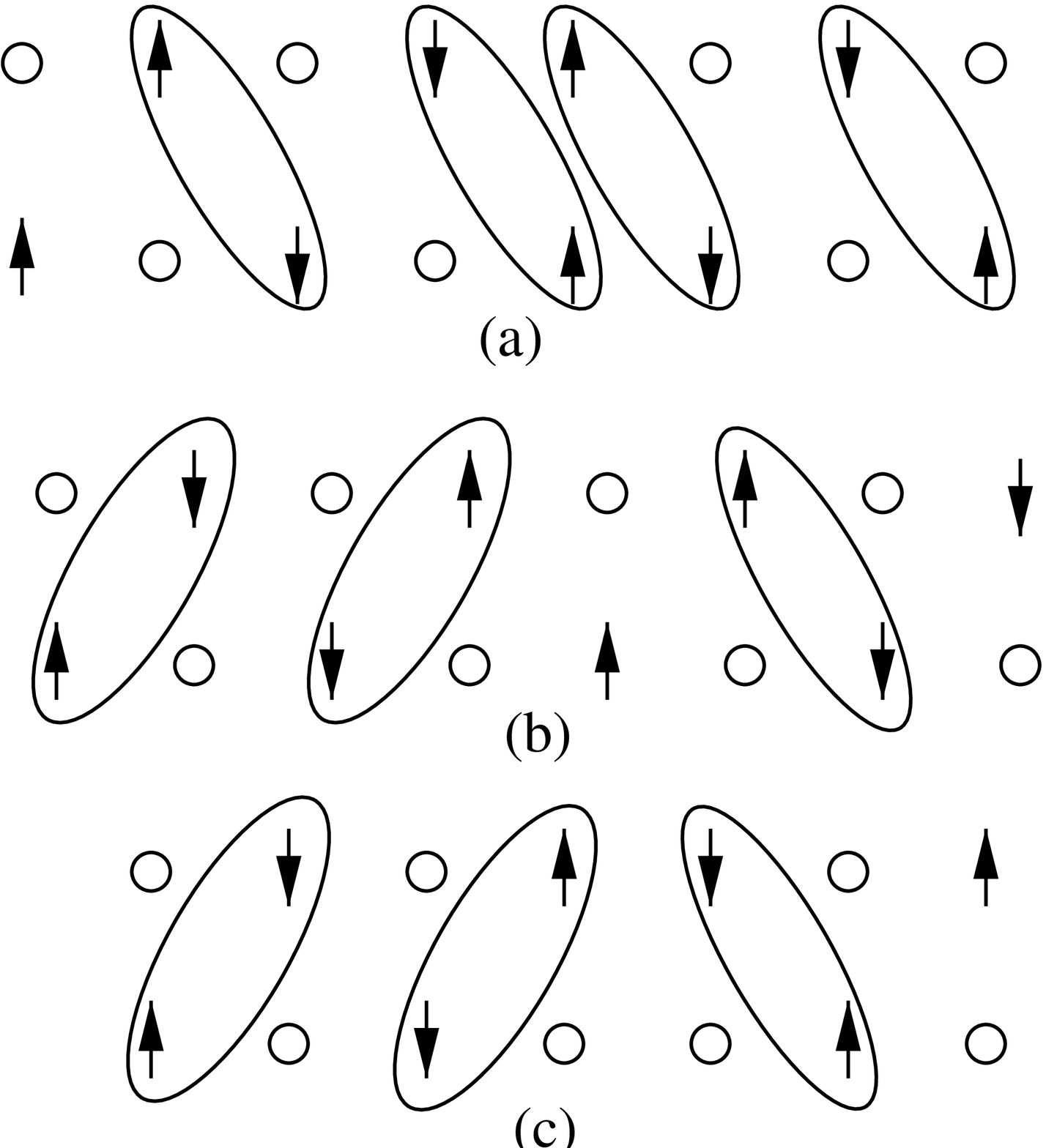}
    \caption{Excitations of the quarter filled ladder in the zig-zag
      charge ordered phase: \\ (a) Charged excitation: it
      can be viewed as a domain wall between the two zig-zag charge
      ordered ground states of the ladder
  or equivalently as a bound state of two
      antiholons of the  quarter-filled chains. \\ (b) spinon
      excitation. It can be viewed as a domain wall
      on the spontaneously dimerized
      effective zig-zag ladder or as spinon inside one of the
      chains. \\ (c) Neutral excitations. It can be viewed either as a
      domain wall between the two zig-zag charge
      ordered ground states or as a holon-antiholon bound state.}
    \label{fig:excit-zig}
  \end{center}
\end{figure}

These solitons can be understood as domain walls between the two charge
ordered ground states as represented
 on figure~\ref{fig:excit-zig}(a)
and figure~\ref{fig:excit-zig}(c).
Alternatively, the charged solitons can be viewed as
holon-holon (for charge $-e$)  or antiholon-antiholon (for charge
$+e$) bound states. The neutral solitons can similarly be viewed
 as holon-antiholon bound state. An immediate consequence of the
 mapping of the charge excitations onto the sine-Gordon model is that
 the transport properties of the quarter-filled ladder can be obtained
 from the methods reviewed in Ref. \cite{controzzi_mott}.
For $K_{\rho\alpha}>1/4$, the solitons are the only possible excitations in the model.
For more strongly repulsive excitations, the formation of soliton bound states
(breathers) becomes possible. The existence of these excitations translates into an
exciton peak in the optical conductivity. Such exciton peak has been studied in the
case of the single Hubbard chain in Ref. \onlinecite{essler_mott_excitons1d}. However,
in the model we are considering, for $K_{\rho \alpha}<1/4$, intrachain Umklapp scattering
becomes relevant and  total/antisymmetric charge excitations no more
decouple in this regime.

\subsubsection{spin excitations}
\label{sec:spin-excitations}
To describe spin excitations, we need to consider the effective Hamiltonian
of the spin modes:
\begin{eqnarray}
  \label{eq:spin-only}
  H=\int \frac{dx}{2\pi} \sum_{r=\pm} \left[ u_{\sigma,r} K_{\sigma,r} (\pi
    \Pi_{\sigma,r})^2 + \frac{u_{\sigma,r}}{ K_{\sigma,r}} (\partial_x
    \phi_{\sigma,r})^2 \right] +\frac{2g_2}{(2\pi\alpha)^2} \int dx
  \cos 2\phi_{\sigma+} \cos 2\theta_{\sigma-}
\end{eqnarray}
This Hamiltonian corresponds exactly to the bosonized Hamiltonian of a zig-zag spin
ladder\cite{white_zigzag,allen,allen_spinons}. Remarkably, in the
strong coupling limit $V_\parallel,V_\perp,U \gg t_\perp,t$,
a mapping on a zig-zag spin chain was derived to describe the low
energy excitations in the $(\pi,\pi)$-CDW phase\cite{vojta_qfl}. We
thus notice the continuity between the weak coupling and the strong
coupling limit for the spin excitations of the $(\pi,\pi)$-CDW in this
problem.
Spin excitations of the quarter filled
ladder are those of a zig-zag ladder,
i. e. massive spinons. A more detailed picture of the spin
excitations can be gained by applying a transformation due to Witten and
Shankar\cite{witten_shankar,allen_spinons,zachar_exotic_kondo}. We first notice that
we have $K_{\sigma+} K_{\sigma-}=1$ and $u_{\sigma+}=u_{\sigma-}$. Thus, we can
perform a duality transformation $\tilde{\phi}_{\sigma-}=\theta_{\sigma-},
\tilde{\theta}_{\sigma-}=\phi_{\sigma-}$, and introduce:
\begin{eqnarray}
  \label{eq:transf}
  \phi_a=\frac{\phi_{\sigma+}+\tilde{\phi}_{\sigma-}}{\sqrt{2}} \\
\phi_b=\frac{\phi_{\sigma+}-\tilde{\phi}_{\sigma-}}{\sqrt{2}}
\end{eqnarray}
This procedure permits us to reduce the effective Hamiltonian to:
\begin{eqnarray}
  \label{eq:decoupled-zigzag}
&&  H=H_a+H_b \\
&& H_a=\int \frac{dx}{2\pi}\left[ u_a K_a (\pi
    \Pi_a)^2 + \frac{u_a}{ K_a} (\partial_x
    \phi_a)^2 \right] -\frac{g_2}{(2\pi\alpha)^2} \int dx \cos
  \sqrt{8} \phi_a \\
&& H_b=\int \frac{dx}{2\pi}\left[ u_b K_b (\pi
    \Pi_b)^2 + \frac{u_b}{ K_b} (\partial_x
    \phi_b)^2 \right] -\frac{g_2}{(2\pi\alpha)^2} \int dx \cos
  \sqrt{8} \phi_b,
\end{eqnarray}
where $u_a=u_b=u_{\sigma+}$, $K_a=K_b=K_{\sigma+}$ and $g_2$ preserves $SU(2)$
symmetry. The original Hamiltonian (\ref{eq:spin-only}) is thus decoupled into two
massive sine-Gordon model at the $SU(2)$ point. They describe two massive spin
excitations carrying spin $1/2$. In the strong coupling limit, the spin excitations
can be viewed as the spinons of a zig-zag ladder (see figure \ref{fig:excit-zig}).

\subsection{$(\frac \pi 2,\frac \pi 2)$-CDW}
In the $(\frac \pi 2,\pm \frac \pi 2)$-charge density wave phases, the
charge excitations carry the same
quantum numbers as in the zig-zag charge ordered phase. They
correspond to domain walls between the four different $(\frac \pi
2,\frac \pi 2)$-CDW ground states.
However, spin excitations are
of a different nature since both $S_1^z$ and $S^z_2$ are
good quantum numbers. In fact, these  spin excitations are  massive spinons ``confined'' within
each chain.

\subsection{phases with $\langle \phi_{\rho-}\rangle=\frac \pi 2$}
\begin{figure}[htbp]
  \begin{center}
    \includegraphics[width=\figwidth]{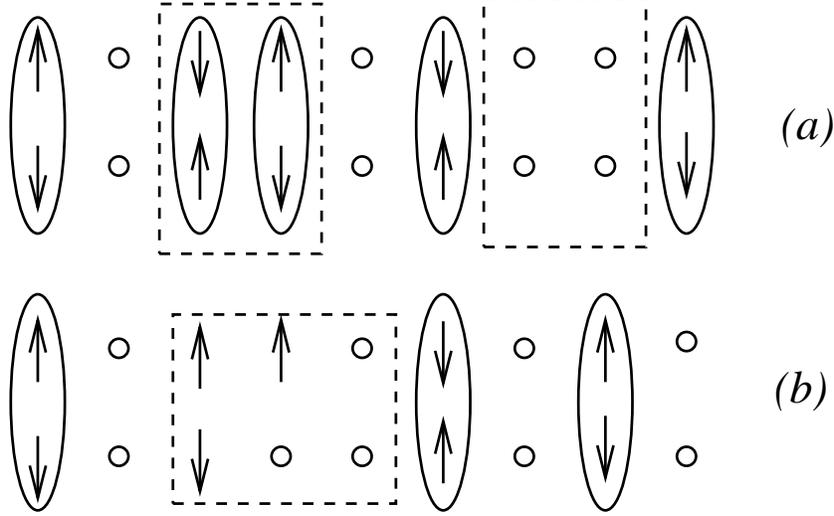}
    \caption{Elementary excitations in the $(\pi,0)$-CDW. (a) holon
      and antiholon excitations carrying the charge $\pm e$
      respectively. (b) Mixed spin/orbital excitation.}
    \label{fig:excit-mixed}
  \end{center}
\end{figure}

The total charge excitations still carry $\pm e$, as in the previous phase, but this
time it is not possible to decouple spin excitations from antisymmetric charge
excitations. When $\phi_{\sigma-}$ is ordered, we have
$\phi_{\rho-}(+\infty)-\phi_{\rho-}(-\infty)=\pm \pi/2$,
$\phi_{\sigma+}(+\infty)-\phi_{\sigma+}(-\infty)=\pm \pi/2$ and
   $\phi_{\sigma-}(+\infty)-\phi_{\sigma-}(-\infty)=\pm \pi/2$.
As a result, the elementary excitations carry total charge zero,
charge difference $\pm e$, and spin $S^z_1=\pm 1/2$ or $S^z_2=\pm
1/2$. These elementary excitations can be viewed as a three body bound
state of a holon in one chain, a antiholon in the other chain and a
spinon. This excitation is sketched on Fig.~\ref{fig:excit-mixed}.

In fact, there exists an interesting analogy between the spin/antisymmetric
charge sector of the quarter filled ladder and the half-filled
Hubbard-Kondo-Heisenberg  (HKH) chain.

The half-filled HKH chain is described by the
Hamiltonian:
\begin{eqnarray}
  \label{eq:kondo}
  H=-t \sum_{i,\sigma} (c^\dagger_{i+1,\sigma} c_{i,\sigma} +
  c^\dagger_{i,\sigma} c_{i+1,\sigma}) + U \sum_i n_{i,\uparrow}
  n_{i,\downarrow} + \frac{J_K}{2}
\sum_{i,\alpha,\beta} \vec{S}_i \cdot
  c^\dagger_{i,\alpha} \vec{\sigma}_{\alpha,\beta} c_{i,\beta} + J_H
  \sum_i \vec{S}_i \cdot \vec{S}_{i+1},
\end{eqnarray}
where $\sigma^{x,y,z}$ are the usual Pauli matrices. At half-filling, the resulting
bosonized Hamiltonian~\cite{fujimoto_kondo1d,lehur98_kondo1d,lehur00_kondo1d} reads:
\begin{eqnarray}
  \label{eq:kondo-bosonized}
H&& =\int \frac{dx}{2\pi} \left[u_\rho K_\rho (\pi \Pi_\rho)^2 +
    \frac{u_\rho}{K_\rho} (\partial_x \phi_\rho)^2\right] +
\int \frac{dx}{2\pi} \left[u_{\sigma+} K_{\sigma+} (\pi \Pi_{\sigma+})^2 +
    \frac{u_{\sigma+}}{K_{\sigma+}} (\partial_x \phi_{\sigma+})^2\right]\\
&& +\int \frac{dx}{2\pi} \left[u_{\sigma-} K_{\sigma-} (\pi \Pi_{\sigma-})^2 +
    \frac{u_{\sigma-}}{K_{\sigma-}} (\partial_x
    \phi_{\sigma-})^2\right]
+ \frac{2J_K}{(2\pi a)^2} \int dx \cos \sqrt{2} \phi_\rho \left[ \cos
  2\phi_{\sigma-} -\cos 2\phi_{\sigma+} + 2 \cos 2
  \theta_{\sigma-}\right].
\end{eqnarray}
Making a rescaling $\phi_\rho = \sqrt{2} \phi_{\rho-}$,$K_\rho = 2K_{\rho-}$, we
obtain the same Hamiltonian as the one describing the spin/charge difference
excitations of the quarter-filled ladder~~(\ref{eq:full})-(\ref{eq:exch-ham}). In
particular, the point $K_{\rho-}=1/2$ in the quarter filled ladder corresponds to
$K_\rho=1$ in the Kondo-Heisenberg chain, i.e. $U=0$ in
(\ref{eq:kondo}).
Thus, magnetic properties near the metal insulator transition of the quarter
filled ladder should be analogous to those of a Kondo-Heisenberg
insulator.
Close to $K_{\rho-}=1$, the structure of excitations is quite different. In the
vicinity of that point\cite{schulz_son}, gapful excitations of the ladder away from
half-filling are described by a massive $SO(3)\times SO(3)$ Gross-Neveu model\cite{azaria_su4}
as a consequence of the breaking of the original
$SU(4)$ symmetry to $SU(2)\times
SU(2)$. The enhancement of the symmetry by the RG is instead absent for
$K_{\rho-}=1/2$.

\section{Numerical study of the phase diagram}
\label{sec:numerics}

In this Section we discuss the numerical results of the RG equations. Prior than that
we analyze in details the nature of the transition among the various phases. The analysis
is partially accomplished by a mapping onto a Majorana fermion theory, that gives also rise
to a connection with the spin-orbital models.

\subsection{Mapping on a theory containing Majorana
  Fermions}\label{sec:majorana}

 We consider first the part of the Hamiltonian (\ref{eq:exch-ham}) with couplings
 $g_{4,5,6}$.
It is convenient to rewrite this part of the interaction
in terms of massive Majorana fermion operators $\xi_\nu$ ($\nu=0,1,2,3$). Using the
identities\cite{shelton_spin_ladders}:
\begin{eqnarray}
  \label{eq:majorana-terms}
  \frac{\cos 2 \phi_{\sigma+}}{\pi a}=i
  (\zeta_{R,1}\zeta_{L,1}+\zeta_{R,2}\zeta_{L,2}),\\
\frac{\cos 2 \phi_{\sigma-}}{\pi a}=i
(\zeta_{R,3}\zeta_{L,3}+\zeta_{R,0}\zeta_{L,0}), \\
\frac{\cos 2 \theta_{\sigma-}}{\pi a}=i
(\zeta_{R,0}\zeta_{L,0}-\zeta_{R,3}\zeta_{L,3}),
\end{eqnarray}
\noindent and the relation $g_5=g_4+g_6$, the interaction terms  $g_{4,5,6}$
are rewritten as:
\begin{eqnarray}
  \label{eq:inter-majo}
  i \cos 2\phi_{\rho-} \left[\frac{g_4}{2\pi
      a}(\zeta_{R,1}\zeta_{L,1}+\zeta_{R,2}\zeta_{L,2}+\zeta_{R,3}\zeta_{L,3})
      +\frac{g_5+g_6}{2\pi a} \zeta_{R,0}\zeta_{L,0}\right],
\end{eqnarray}
where $g_5+g_6$ is always a positive quantity.

In the continuum limit, following
Ref.\onlinecite{shelton_spin_ladders} we can express the bosonic exponents in terms of
the order ($\sigma_i$)  and
disorder ($\mu_i$) parameters of four Ising models as follows:

\begin{eqnarray}
  \cos \phi_{\sigma+}=\sigma_1 \sigma_2,\\
  \cos \phi_{\sigma-}=\sigma_3 \sigma_0,\\
  \sin \phi_{\sigma+}=\mu_1 \mu_2,\\
  \cos \theta_{\sigma-}=\mu_3 \sigma_0.
\end{eqnarray}

When $\langle\phi_{\rho-}\rangle=\frac \pi 2$, the Ising order parameter $\sigma_0$
has a non-zero expectation value. However the sign of $g_4$ is not fixed. For $g_4>0$,
and $\langle\phi_{\rho-}\rangle=\frac \pi 2$, $\sigma_{1,2,3}$ have all nonzero
expectation values, corresponding to have both $\phi_{\sigma+}$ and $\phi_{\sigma-}$
ordered. For $g_4<0$, $\mu_{1,2,3}$ have all nonzero expectation values, corresponding
to having both $\phi_{\sigma+}$ and $\theta_{\sigma-}$ ordered.

Now, we would like to focus on the interaction part of the Hamiltonian that contains
the terms $g_1,g_2,g_{\sigma\pm}$. Using the mapping on Majorana fermions, we can
write this part as:
\begin{eqnarray}
\label{major_g12s}
&& \frac{g_{\sigma+}}{2\pi^2}(\partial_x \phi_{\sigma+})^2 +
  \frac{g_{\sigma-}}{2\pi^2}(\partial_x \phi_{\sigma+})^2 +
  \frac {g_1}{2(\pi a)^2} \cos 2 \phi_{\sigma-} \cos 2 \phi_{\sigma+}
+ \frac {g_2}{2(\pi a)^2} \cos 2 \theta_{\sigma-} \cos 2
\phi_{\sigma+} \\
&& = - g_{\sigma+} \zeta_{R,1} \zeta_{L,1} \zeta_{R,2}
\zeta_{L,2}- g_{\sigma-} \zeta_{R,0} \zeta_{L,0} \zeta_{R,3}
\zeta_{L,3} - g_+ (\zeta_{R,1} \zeta_{L,1}+ \zeta_{R,2}
\zeta_{L,2})\zeta_{R,0} \zeta_{L,0} -g_- (\zeta_{R,1} \zeta_{L,1}+
\zeta_{R,2}\zeta_{L,2})\zeta_{R,3} \zeta_{L,3}.\nonumber 
\end{eqnarray}
\noindent where $g_\pm=(g_1\pm g_2)/2$. Using the $SU(2)$ symmetry conditions(\ref{eq:hypersurface}), this
expression can be further rewritten as:
\begin{eqnarray}
  - g_{\sigma+} (\zeta_{R,1} \zeta_{L,1} \zeta_{R,2}
\zeta_{L,2} + \zeta_{R,1} \zeta_{L,1} \zeta_{R,3}
\zeta_{L,3} +\zeta_{R,2} \zeta_{L,2} \zeta_{R,3}
\zeta_{L,3}) - g_{\sigma-}  (\zeta_{R,1} \zeta_{L,1}+
\zeta_{R,2}\zeta_{L,2}+  \zeta_{R,3} \zeta_{L,3})  \zeta_{R,0} \zeta_{L,0},
\end{eqnarray}
\noindent which makes the $SU(2)$ symmetry transparent. We note that for
$g_{\sigma-}>0$, $i\langle \zeta_{R,a} \zeta_{L,a}\rangle$ can have opposite signs
depending whether $a=0$ or $a\ne 0$. This implies that either $\langle\mu_0
\rangle\ne0$ and $\langle \sigma_{1,2,3}\rangle \ne 0$ or $\langle\sigma_0
\rangle\ne0$ and $\langle \mu_{1,2,3}\rangle \ne 0$. Both cases correspond to having
$\theta_{\sigma-}$ and $\phi_{\sigma+}$ ordered. When $g_{\sigma-}<0$,
we have both $\phi_{\sigma+}$ and $\phi_{\sigma-}$ ordered.

\subsubsection{Spin-orbital models}

Before closing this section, we would like to discuss the following connection. If we
consider the case of $V_\parallel=0$, $J_\perp=0$ and $V_\perp=U$, we can rewrite the
interchain interaction in Eq.~(\ref{eq:latt-hubbard}) as:
\begin{equation}
\label{eq:su4}
  \frac U 2 \sum_i
  (n_{i,1,\uparrow}+n_{i,1,\downarrow}+n_{i,2,\uparrow}+n_{i,2,\downarrow})^2,
\end{equation}
\noindent which has a manifest $SU(4)$ invariance. Thus the problem is
related to the
quarter-filled $SU(4)$ Hubbard model\cite{assaraf_su(n)}. The charge
Umklapp term
derived for that model\cite{assaraf_su(n)} agrees with
(\ref{eq:charge-hamiltonian}). If U is large enough, the low energy
excitations are those of an antiferromagnetic chain of $SU(4)$
spins\cite{assaraf_su(n)}. Considering deviations from the $SU(4)$
symmetric point, it is possible to derive a model describing the low
energy excitations of the insulator\cite{mostovoy_nav2o5} in terms of coupled spin and
orbital modes (the spin-orbital model). In
Refs.~\onlinecite{azaria_su4,azaria_su4_long,itoi_spin_orbital}, a
 $SU(2)\times SU(2)$ spin-orbital symmetric model
was analyzed perturbatively around the $SU(4)$ point using bosonization and
refermionization techniques. An Hamiltonian describing the low energy
dynamics of the system with six Majorana fermions was obtained, and the
formation of a dimerized spin gapped phase was predicted in the
physical range of parameters. The existence of this dimerized phase
was also confirmed by numerical
studies\cite{itoi_spin_orbital,pati_orbital_dmrg}.
In our
present problem, assuming $K_{\rho-}\sim 1$ so that the term
containing $g_3$ can be neglected, and using the mapping:
\begin{equation}
  \label{eq:phirhomoins-orbital}
  \frac{\cos 2 \phi_{\rho-}}{\pi a}=i
  (\zeta_{R,4}\zeta_{L,4}+\zeta_{R,4}\zeta_{L,4}),
\end{equation}
we can recast the interactions in the spin/antisymmetric charge
sector of Eq.~(\ref{eq:exch-ham}) in terms of two triplets of Majorana
fermions: $(\zeta_1,\zeta_2,\zeta_3)$ representing the spin
excitations, and $(\zeta_0,\zeta_4,\zeta_5)$ representing the orbital
excitations as in
Ref.~\onlinecite{azaria_su4,azaria_su4_long,itoi_spin_orbital}. The
Hamiltonian we obtain is however more general, since we did not assume
any $SU(2)$ symmetry in the orbital sector. From our previous
discussion of the phase diagram in Sec.~\ref{sec:phase} we expect that
the dimerized phase (Phase IV in Ref.~\onlinecite{itoi_spin_orbital}
or Phase A in Ref.~\onlinecite{azaria_su4_long}) of the spin-orbital
model is related to the $(\pi,0)$-BOW. In a different
limit\cite{nersesyan_biquadratic,orignac_spintube} of the
$SU(2)\times SU(2)$ spin-orbital
model, a dimerized insulating phase analogous to the $(\pi,0)$-BOW was
also obtained.  We note that no charge ordering is predicted by the
$SU(2)\times SU(2)$ spin orbital model. The reason  for this
is that the $V_\parallel$ term of the ladder model
produces a large renormalization of
$K_{\rho-}$ in the charge ordered state, while leaving $K_{\sigma-},K_{\sigma+}$ close
to the non-interacting value, due to the remaining $SU(2)$ symmetry.
To describe charge ordering, one has thus to consider more general
spin-orbital models in which the interaction in the orbital sector has only the
$U(1)$ symmetry\cite{mostovoy_nav2o5}.

\subsection{order of the transitions between the different phases}
We now turn to the quantum phase transitions between the different
phases of the quarter-filled ladder. These phase transitions can
result from a change of the ordering in the $\phi_{\rho-}$ field, the
spin gap being preserved, or from a change of the ordering in the spin
sector. As we will see below, the former phase transitions are of
second order and belong to the one-dimensional quantum Ising
universality class,  whereas the latter transitions can be either of
second and first order. We begin discussing the Ising
transitions in the antisymmetric charge mode.
\subsubsection{Ising transitions}
We consider  phase transitions in which the order in the spin sector
is not changed, but the order in the antisymmetric charge sector
$\phi_{\rho-}$ is modified.
These transitions  occur between the $(\pi,0)$-BOW and the
$(\pi,\pi)$-CDW, the $(\frac \pi 2, \pi)$-CDWs and the $(\frac \pi
2,\frac \pi 2)$-CDWs, the $(\pi,\pi)$-BOW and the $(\pi,0)$-CDW.
Since the order in the spin sector does not change we can describe the
transition by concentrating only on the $\phi_{\rho-}$ modes. The
resulting effective Hamiltonian reads:
\begin{equation}
  \label{eq:ham-rho-minus}
  H_{\rho-}=\int \frac{dx}{2\pi} \left[u_{\rho-}K_{\rho-}(\pi
    \Pi_{\rho-})^2 + \frac {u_{\rho-}}{K_{\rho-}} (\partial_x
    \phi_{\rho-})^2\right] +\frac{2g_3}{(2\pi a)^2} \int dx \cos
  4\phi_{\rho-} +\frac{2\overline{g}}{(2\pi a)^2} \int dx \cos
  2\phi_{\rho-}.
\end{equation}
\noindent This Hamiltonian is the one of the double sine-Gordon
model\cite{fabrizio_dsg}. The semiclassical analysis of the double
cosine potential shows that for $\langle \phi_{\rho-}\rangle= \pm
\frac \pi 4$, $\langle \cos 2\phi_{\rho-}\rangle=0$.
Using the result of \cite{fabrizio_dsg}, this implies that as
$\overline{g}$ is varied, an Ising transition is obtained in
$\phi_{\rho-}$. For the sake of simplicity, let us consider the
transition between the $(\pi,0)$-BOW and the $(\pi,\pi)-$CDW. Similar
features appear in all the other Ising transitions. In the
$(\pi,\pi)$-CDW, the order parameter $O_{CDW_{(\pi,\pi)}}\sim \sin
  2\phi_{\rho-}$. Using the results of \cite{fabrizio_dsg} on the
  ultraviolet-infrared transmutation of operators, it is easily seen that
$O_{CDW_{(\pi,\pi)}}\sim \mu$, where $\mu$ is the disorder parameter of
  a quantum Ising model. This implies in particular that
  $\langle O_{CDW_{(\pi,\pi)}}(x)O_{CDW_{(\pi,\pi)}}(0)\rangle \sim
  x^{-1/4}$ at the transition. Moreover,
  near the transition one has: $\langle O_{CDW_{(\pi,\pi)}}\rangle
  \sim \left(\frac {a \Delta_{\rho-}}{v}\right)^{1/8}$. If the gap  $\Delta_{\rho-}$
  vanishes linearly with the interaction $V$ this
  implies $\langle O_{CDW_{(\pi,\pi)}}\rangle
  \sim (V-V_c)^{1/8}$, giving rise to the onset of the order parameter at a critical value
  $V_c$. Such onset was numerically observed at the
  homogeneous insulator HI$_{sg}$-CDW$_{sg}$ transition in Ref.\onlinecite{vojta_qfl} and
  attributed to an Ising transition.
In our results, we observe that neither the charge gap nor the
spin gap go to zero at the transition, meaning the absence of a critical value $V_c$
for the onset of the transition.

A simple picture of the transition
is obtained, in the limit of strong coupling $V_\perp \to \infty$,
by an effective
quantum Ising model. In this picture, the only states retained are the
electron pairs forming singlets along the diagonal and pointing either
in the northwest or northeast direction.
The variable
$\sigma^z=1$ when the diagonal singlet formed of two electrons is
oriented northwest (NW), and $\sigma^z=-1$ when the singlet is oriented northeast (NE). In the
ground state all the electron pairs must have the same orientation.
The corresponding potential energy reads:
\begin{equation}
  \label{eq:q-ising-pot}
  H_{pot.}=-V_\parallel \sum_n \sigma_n^z\sigma_{n+1}^z.
\end{equation}
\noindent The kinetic energy comes from the term $t$. In
second order perturbation theory, $t$ flips a singlet pair
from the NW to NE orientation. This process leads to a
kinetic term in the Hamiltonian:
\begin{equation}
\label{eq:q-ising-kin}
 H_{kin.}=\frac{t_\parallel^2}{V_\perp} \sigma_n^x.
\end{equation}
\noindent When the kinetic term in the Hamiltonian dominates, the
singlet pairs go back and forth between NE and NW orientation leading to a
zero average of the $(\pi,\pi)$-CDW order parameter, and an effective
bond order wave. When the potential energy dominates, the singlet
pairs are all locked in the NE or NW orientation giving rise to a
nonzero $(\pi,\pi)$-CDW.

\subsubsection{spin transitions}
In the present section, we discuss the transitions in the spin sector.
Since the gaps in the$\phi_{\rho-}$ and $\phi_{\rho+}$ sectors are
robust, we can concentrate on a low energy effective spin model.
First, let us focus on the case of $\langle \phi_{\rho-}\rangle=\frac \pi
2$. In that case, the discussion is equivalent to the one in
Ref.~\onlinecite{tsuchiizu_2leg_firstorder}. The theory describing the
transition point is the $O(3)$ Gross-Neveu model, and the operator
that takes the system away from the transition point is the mass of the
Gross-Neveu fermions. As a result, the system will have a
second order phase transition in the $SU(2)_2$ WZW model
universality class when the $O(3)$ Gross-Neveu has no spontaneous
symmetry breaking and a first order transition when the Gross-Neveu
model presents dynamical mass
generation\cite{shankar_gn_at,goldschmidt_susy}. For $g_{\sigma+}<0$,
there is a spontaneous symmetry breaking and thus a first order
transition. Since this corresponds to $J_\perp>0$, first order spin
transitions should be generic in the models we are considering.
In particular, first order transitions should be expected between
phases such as the $(\pi,0)$-BOW and the $(\frac\pi 2,\pi)$-CDW or the
$(\pi,0)$-CDW and the $(\frac \pi 2,\pi)$-CDW. These first order
transitions occur in the spin sector and should be observable by
looking at spin-spin correlations.
In the case of $\langle \phi_{\rho-}\rangle=\frac \pi 4$, we have to focus on
the terms coming from $g_1,g_2,g_{\sigma\pm}$. In that case, for a
transition to be possible we must have
$g_{\sigma-}=0$ as one can see directly from Eq.~(\ref{major_g12s}).
The behavior at the transition then depends on the sign of $g_{\sigma+}$. 
When $g_{\sigma+}>0$, no gap is generated in the triplet modes thus giving a $SU(2)_1\times
SU(2)_1$ criticality. For $g_{\sigma+}<0$, the triplet modes remain massive at the transition, leading to an Ising criticality. Since $g_{\sigma+}\sim -J_\perp<0$,  an Ising  transition is obtained between
the $(\pi,\pi)$-BOW and the $(\frac \pi 2,\frac \pi 2)$-CDW or the
$(\pi,\pi)$-CDW and the $(\frac \pi 2,\frac \pi 2)$-CDW.

\subsection{RG calculation}
\begin{figure}[htbp]
  \begin{center}
    \includegraphics[width=\figwidth]{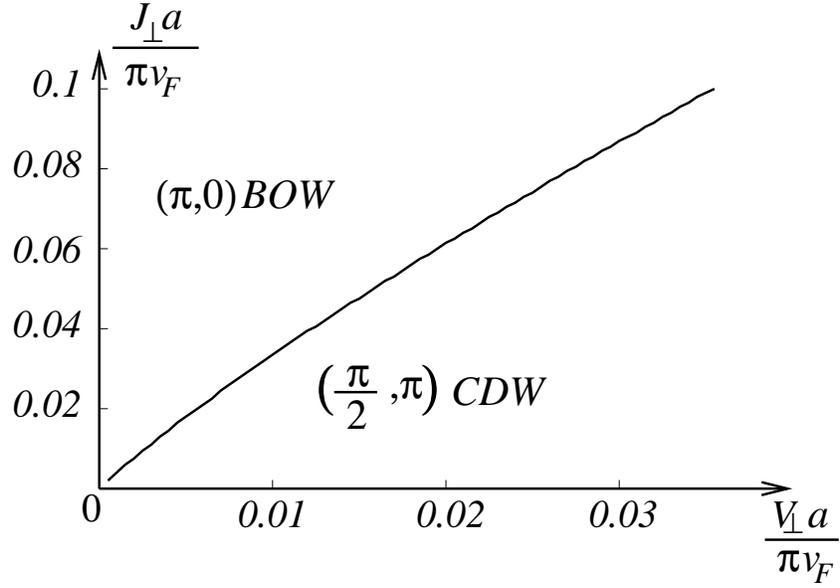}
    \caption{The topology of the phase diagram of the quarter filled
      ladder for $K_{\rho-}\alt 1/2$. Only the $(\pi,0)$-BOW and the
      $(\frac \pi 2,\pi)$-CDW are obtained.}
    \label{fig:phasediag1}
  \end{center}
\end{figure}

\begin{figure}[htbp]
  \begin{center}
    \includegraphics[width=\figwidth]{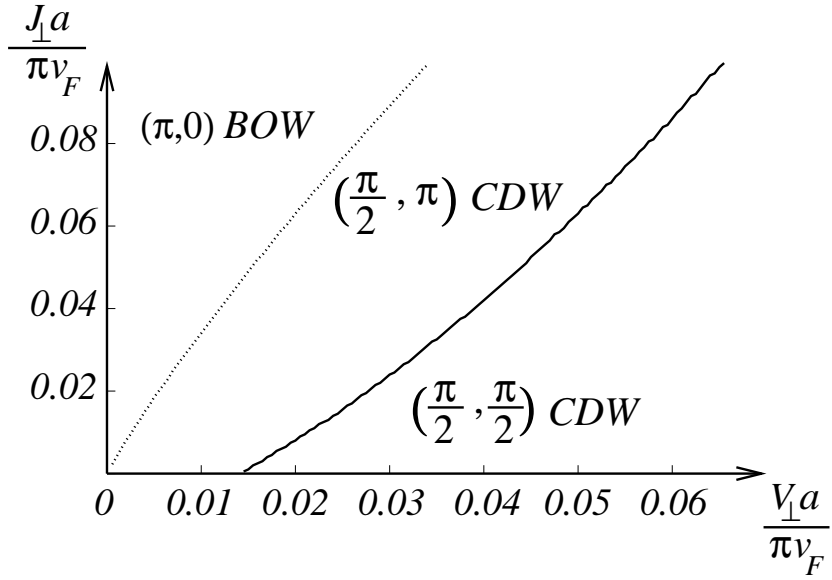}
    \caption{The topology of the phase diagram of the quarter filled
      ladder for $1/2>K_{\rho-}\agt 1/3$. The $(\frac \pi 2,\frac \pi
      2)$-CDW appears at large $V_\perp$.}
  \label{fig:phasediag2}
  \end{center}
\end{figure}

\begin{figure}[htbp]
  \begin{center}
 \includegraphics[width=\figwidth]{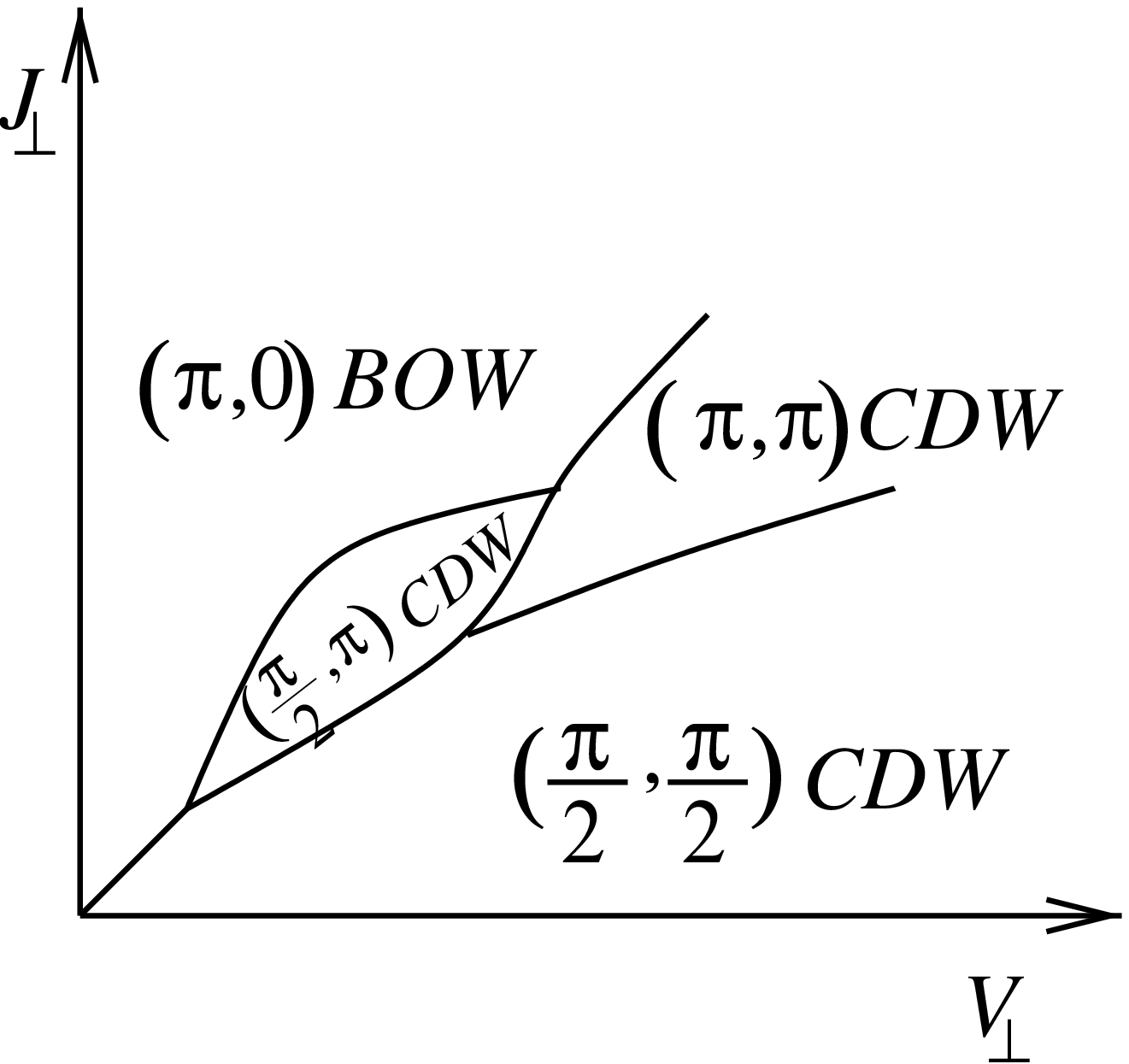}
    \caption{The topology of the phase diagram of the quarter filled
      ladder for $K_{\rho-}\alt 1/3$. The size of the $(\frac \pi
      2,\pi)$-CDW phase region has been exaggerated.}
  \label{fig:phasediag3}
  \end{center}
\end{figure}

\begin{figure}[htbp]
  \begin{center}
    \includegraphics[width=\figwidth]{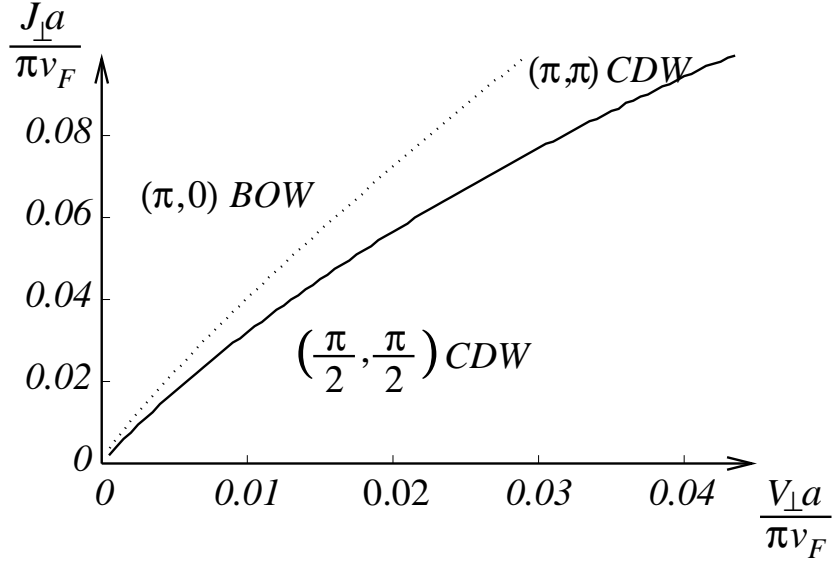}
    \caption{The topology of the phase diagram of the quarter filled
      ladder for $K_{\rho-}< 1/3$. The $(\pi,0)$-BOW is obtained for
      strong $J_\perp$ and the $(\frac \pi 2,\frac \pi 2)$-CDW is
      obtained for strong $V_\perp$. The $(\pi,\pi)$-CDW is obtained
      in the intermediate regime.}
\label{fig:phasediag4}
  \end{center}
\end{figure}

To find the phase diagram, we integrate the Eqs.~(\ref{rgred}) numerically using a
fourth order Runge-Kutta algorithm for fixed values of $K_{\rho-}$ at varying
$V_\perp$ and $J_\perp$. We stop the numerical integration when one of the coupling
constants $y_3,y_5,y_6$ becomes of order $1$.
We find that at this scale, $y_1,y_2$
are still inferior to $1$.
We have the following results. First, for $K_{\rho-}$
larger than $1/3$, ($K_{\rho-}\simeq 1/2$) the $(\frac \pi 2,\frac \pi
2)$-CDW and $(\pi,\pi)$-CDW are absent. This is a consequence of the fact that in
this regime, the $2k_F$ fluctuations are dominant over the $4k_F$
ones. As a result, we find either
 the bond order wave $(\pi,0)$-BOW or the $(\frac \pi 2,\pi)$-CDW. As could be expected, a large
$J_\perp$ favors the former, and a large $V_\perp$ favors the latter.
A first order transition is expected between these two phases.
The  phase
diagram for $K_{\rho-}=0.5$  is drawn on Fig.~\ref{fig:phasediag1}.
When $K_{\rho-}$ becomes smaller  but still larger than $1/3$, the term $\cos
4\phi_{\rho-}$ is more relevant and the  $(\frac \pi 2,\frac \pi
2)$-CDW phase becomes stable at large $V_\perp$. This can be
understood as resulting from an increase in the strength of $4k_F$
fluctuations.  The   phase
diagram for $K_\rho=0.35$  is sketched on Fig.~\ref{fig:phasediag2}.
For $K_{\rho-}$ close to $1/3$, the topology of the phase diagram
becomes more complex. The $(\pi,\pi)$-CDW phase is present for competing
$V_\perp,J_\perp$, along with the $(\pi,0)$-BOW and the two other
CDWs. The phase diagram
is sketched on Fig.~\ref{fig:phasediag3}.
Finally, for $K_{\rho-}<1/3$, $\cos 4\phi_{\rho-}$ is  the most
relevant operator, and the $(\frac \pi 2,\pi)$-CDW
phase disappears altogether. The $(\pi,\pi)$-CDW
phase is obtained for $V_\perp,J_\perp$ intermediate, whereas for
strong $J_\perp$, the $(\pi,0)$-BOW is obtained and for strong $V_\perp$, the
$(\frac \pi 2,\frac \pi 2)$-CDW is obtained.
The phase diagram for $K_\rho-=0.3$ is sketched on
Fig.~\ref{fig:phasediag4}.

\subsection{Commensurate-Incommensurate transitions}
\subsubsection{Deviations from quarter filling}
Away from quarter filling, the number of particles is fixed via a
chemical potential that affects  only $H_{\rho+}$. The charge
modes are described by the Hamiltonian:
\begin{eqnarray}
  \label{eq:doping-ham}
  H_{\rho+}=\int \frac{dx}{2\pi}\left[ u_{\rho+} K_{\rho+} (\pi
    \Pi_{\rho+})^2 + \frac{u_{\rho+}}{K_{\rho+}} (\partial_x
    \phi_{\rho+})^2 \right] -\frac{2\mu}{\pi} \partial_x \phi_{\rho+}
-\frac{2V_\perp a}{(2\pi a)^2} \int dx
  \cos 4\phi_{\rho+}
\end{eqnarray}
where $\mu$ measures the difference in chemical potential. A standard
argument
\cite{japaridze_cic_transition,pokrovsky_talapov_prl,schulz_cic2d}
then shows that for a difference in
chemical potential of the order of the gap of the quarter filled
system, a commensurate-incommensurate (C-I) transition occurs and the system
becomes gapless.  The renormalized $K_{\rho+}^*$
exponent at the gapless point is such\cite{schulz_cic2d}
that the dimension of the operator $\cos
4\phi_{\rho+}$ is one, which gives $K_{\rho+}^*=1/4$.
This result has been previously derived from different arguments
in Ref.~\onlinecite{white_ladder_friedel} and checked against
numerical simulations.

In the incommensurate phase, the $\frac \pi 2$-CDW correlations
are the dominant ones as they behave as $\langle
e^{i\phi_{\rho+}(x)} e^{-i\phi_{\rho+}(0)}\rangle \sim
x^{-K_{\rho+}/2}$. Subdominant
  $\pi$-BOW or $\pi$-CDW correlations are also present and behave as:
 $\langle e^{i2\phi_{\rho+}(x)}
e^{-i2\phi_{\rho+}(0)}\rangle \sim x^{-2K_{\rho+}}$. Right at the
transition, the respective exponents are $1/8$ and $1/2$.

In principle, superconducting correlations are also possible. The only
surviving
superconducting order parameter is the $d$-wave one. Its lattice
expression is given by:

\begin{equation}
O^d_{SC}(i)=\sum_\sigma (c_{i,1,\sigma}c_{i,2,-\sigma}-c_{i,2,\sigma}c_{i,1,-\sigma}).
\end{equation}

One finds that this pairing operator behaves as
$e^{i\theta_{\rho+}(x)}
e^{-i\theta_{\rho+}(0)}\sim x^{-1/(2K_{\rho+})}$. Right at the C-I
transition, these
correlations have an exponent $2$ which indicates that
superconductivity is largely
dominated by CDW correlations. In a doped quarter filled insulator the
presence of the
spin gap is therefore insufficient to render superconducting
correlations dominant. This is to be contrasted with the case of the
half-filled ladder\cite{schulz_mitwochain} where superconducting
correlations are dominant in the conducting phase.

\subsubsection{Effect of a magnetic field}
An applied magnetic field couples to the ladder via a term:
\begin{equation}
  \label{eq:magfield}
  H_{\text{mag.}}=-\frac h \pi \int dx \partial_x \phi_{\sigma+}
\end{equation}
It is well known that a  magnetic field applied to a spin
gap system can produce a Luttinger liquid like
phase\cite{chitra_spinchains_field} when the field becomes larger than
the gap.    
The effect of the magnetic field is different in the case of strong
repulsion ($\langle \phi_{\rho-}\rangle=\frac \pi 4$) and in the case
of weaker repulsion ($\langle \phi_{\rho-}\rangle=\frac \pi 2$).
For strong repulsion the spin-gap is caused by the terms $g_1$ or
$g_2$. The effect of the applied field is to render these two terms
irrelevant. As a result, both $\phi_{\sigma+}$ and $\phi_{\sigma-}$
have gapless excitations leading to a C0S2 phase in the notations of
Ref.\onlinecite{balents_2ch}. This C0S2 phase contains  
in chain $2k_F$ charge density and spin-density wave power-law
correlations. One finds in the case of C0S2 phase obtained by applying a
magnetic field to the $(\pi,\pi)-$CDW or the $(\pi,\pi)-$BOW:
\begin{eqnarray}
  \label{eq:c0s2-exp-zz}
  &&\langle O_{CDW_{2k_F,p}} (x) O_{CDW_{2k_F,p}}(0)
  \rangle \sim  \cos
  (\frac{\pi x} {2a}) \cos (mx) \frac{\text{Const.}}{x^{5/4}},\\
  &&\langle S_p^z(x) S_p^z(0)\rangle \sim \frac 1 {x^2} + \cos
  (\frac{\pi x} {2a}) \cos (mx) \frac{\text{Const.}}{x^{5/4}}, \\
 &&\langle S_p^x(x) S_p^x(0)\rangle \sim \text{Const.} \frac {\cos (mx)}
 {x^{5/2}} + \cos (\frac{\pi x} {2a})\frac{\text{Const.}}{x^{5/4}},
\end{eqnarray}
\noindent where exponents have been obtained from the transformation of
Ref.\cite{allen_spinons} and Sec.~\ref{sec:spin-excitations} by
requiring that at the transition the dimension of the relevant
operators leading to the spin gap in Eq.~(\ref{eq:decoupled-zigzag})
be of dimension one\cite{schulz_cic2d}.
In the case of the C0S2 phase obtained by applying a magnetic field to
the $(\frac \pi 2,\frac \pi 2)$-CDWs, one finds
instead\cite{chitra_spinchains_field}:
\begin{eqnarray}
\label{eq:c0s2-exp-zz-b}
  &&\langle O_{CDW_{2k_F,p}} (x) O_{CDW_{2k_F,p}}(0)
  \rangle \sim  \cos
  (\frac{\pi x} {2a}) \cos (mx) \frac{\text{Const.}}{x^{1/2}}\\
  &&\langle S_p^z(x) S_p^z(0)\rangle \sim \frac 1 {x^2} + \cos
  (\frac{\pi x} {2a}) \cos (mx) \frac{\text{Const.}}{x^{1/2}} \\
 &&\langle S_p^x(x) S_p^x(0)\rangle \sim \text{Const.} \frac {\cos (mx)}
 {x^{5/2}} + \cos (\frac{\pi x} {2a})\frac{\text{Const.}}{x^2}.
\end{eqnarray}
For weaker repulsion, the spin gap is caused by the terms
$g_4,g_5,g_6$. The application of the magnetic field then produces
only a suppression of the gap in $\phi_{\sigma+}$ giving a C0S1
phase. The behavior of the induced spin density wave correlations
depends on whether $\phi_{\sigma-}$ or $\theta_{\sigma-}$ is ordered
in the parent case. In the first case, which corresponds to the
$(\frac \pi 2,\pi)$ CDWs under strong magnetic field, critical correlations develop
in $S^z_p(x)$  as well as in the $(\frac \pi
2,q_y)$-CDW order parameters with an exponent of $1$. In the second
case, which corresponds to the $(\pi,0)$-BOW and the $(\pi,0)$-CDW
under strong magnetic field, only
$S_p^{x,y}$ become critical with an exponent of $1/4$.
\section{Conclusions}

In the present paper, we have studied charge ordering in the
two-leg Hubbard ladder at quarter filling.
Focusing in the regime of strong-coupling on-site Coulomb repulsion $U$,
we have investigated the interplay of the interchain Coulomb repulsion $V_\perp$ and
the exchange interaction $J_\perp$ on charge ordering, and a variety
of spin-gapped charge density waves and bond-order waves have been obtained. In
particular, when the intrachain repulsion is strong enough,
the ground state of the system exhibits a zig-zag
charge order state similar to the phase described in numerical
studies\cite{vojta_qfl}. We have obtained the complete
phase diagram in the $J_\perp-V_\perp$ plane by numerical integration
of perturbative renormalization group
equations and discussed the transitions between the various charge
ordered and bond-ordered states.
The results show that phase transitions can occur by an ordering in the
antisymmetric charge sector or the spin sectors.
The quantum phase transition in the spin sector, as in the half-filled
case, is described by
the $O(3)$ Gross-Neveu
model \cite{tsuchiizu_2leg_firstorder,wu_2leg_firstorder} with a mass
term and can be either second or first order. The transition in the
antisymmetric charge sector which is proper to the quarter-filled
ladder belongs to the Ising universality class. This type of Ising
transition is expected to separate the zig-zag charge ordered state
from a bond ordered wave phase.
We have further analyzed the charge and spin
excitations in the various gapped phases.
Due to charge-spin separation, all of the
insulating phases have spinless holon excitations of charge $\pm e$.
However, the magnetic excitations depend on the nature of
the phases considered.
For strong intrachain repulsion, they can be either massive spinons confined in each
chain in the case of the $(\frac \pi 2, \frac \pi 2)$-CDWs or domain
walls of a dimerized effective zig-zag ladder in the case of a zig-zag
charge ordered state in agreement with Ref.~\onlinecite{vojta_qfl}.
In the case of weaker intrachain repulsion, we have discussed
the analogy of the excitations
spin/antisymmetric charge sector of the quarter filled
ladder with those of a half-filled Hubbard-Kondo-Heisenberg chain.
In the framework of bosonization we have discussed
the connection with other effective models of the quarter-filled ladder,
such as the spin-orbital models\cite{azaria_su4,itoi_spin_orbital}.
We have briefly discussed
the physics away
from quarter filling, where commensurate-incommensurate transitions can occur.
The analysis of correlations functions show that CDW correlations
largely dominate superconducting fluctuations at odds with half-filled
ladders\cite{schulz_mitwochain}. Finally, we have discussed the effect
of magnetic fields strong enough to lift the spin gaps and show that
the induced spin density wave correlations sharply distinguish the
different charge ordered and bond ordered phases.

We would like to point out that although our results do not directly apply
to the $\mathrm{NaV_2O_5}$ compound, since in this material
$t_\perp=2t_\parallel$, our model is able to reproduce a
zig-zag charge ordered state with spin gap. It would be interesting to
compare the features of the zig-zag state we predict with the one obtained in
$\mathrm{NaV_2O_5}$.
In this perspective, it would be interesting to study in details a
model in which interchain hopping is fully taken into account
following the approach of \cite{schulz_2chains,balents_2ch}.
It would also be interesting to extend our analysis to zig-zag ladder models,
since experimental realizations of these
systems are now available\cite{amasaki_2ch_zigzag}.

\begin{acknowledgments}
We thank N. Andrei, M. Cuoco, P. Lecheminant and T. Giamarchi for
discussions and comments. E. Orignac acknowledges support from
Minist\`ere de la recherche et des nouvelles technologies under a grant 
``ACI Jeunes chercheurs''.
\end{acknowledgments}

\appendix

\section{Derivation of the Umklapp term from the lattice
  Hamiltonian of the quarter-filled ladder}\label{app:deriv-umkl-term}

\begin{figure}[htbp]
  \begin{center}
    \includegraphics[width=4cm]{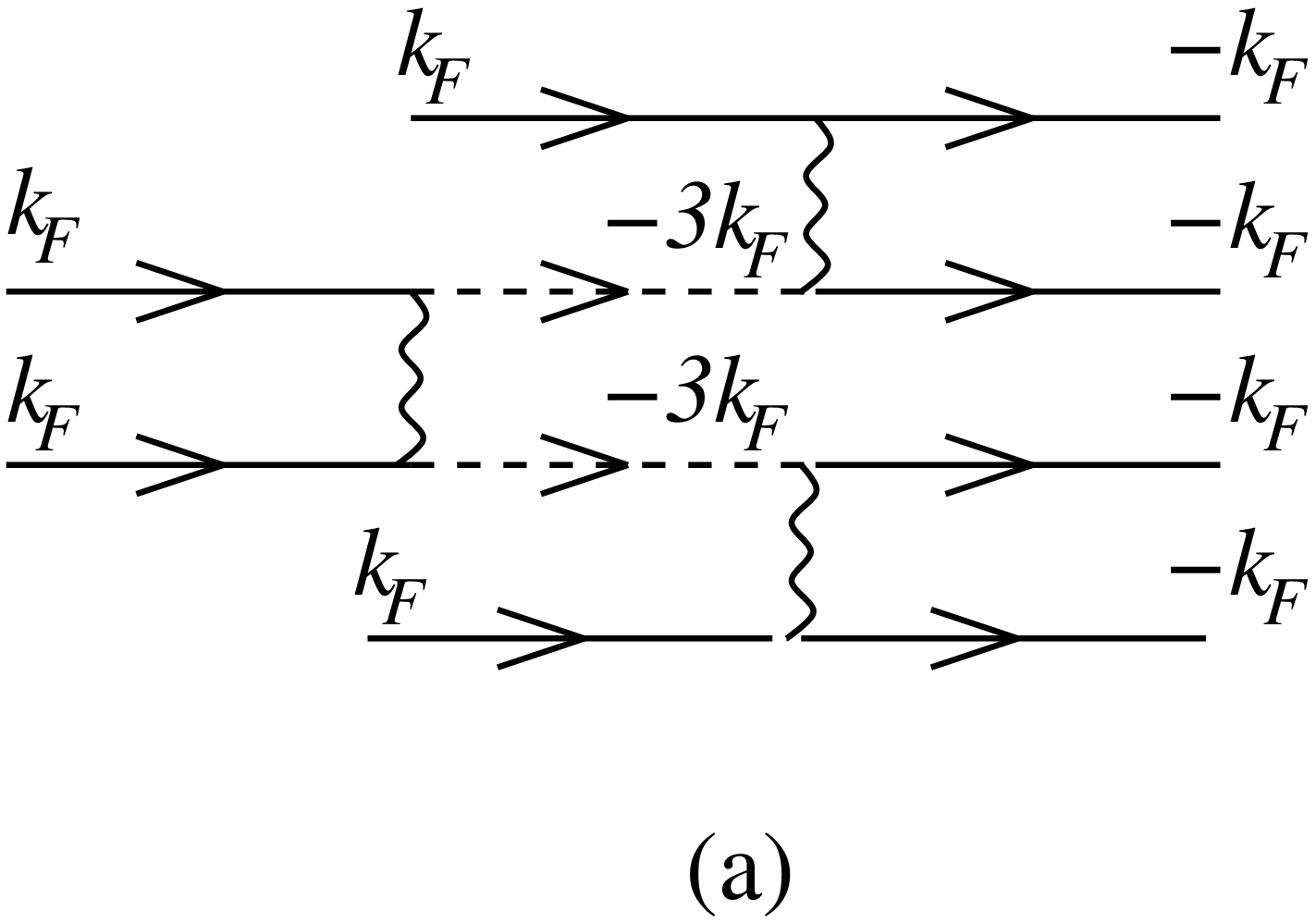}
    \includegraphics[width=6cm]{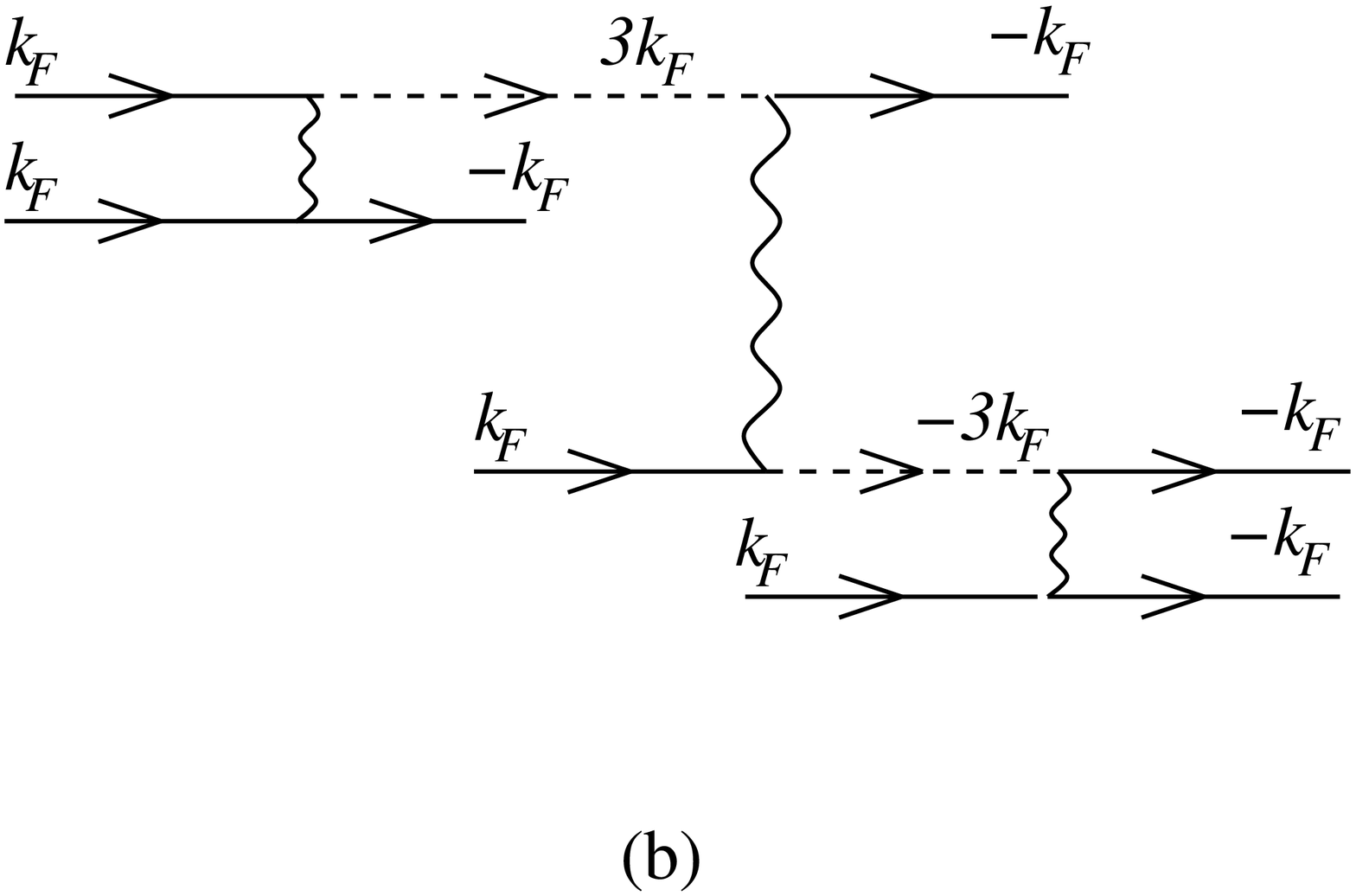}
     \includegraphics[width=6cm]{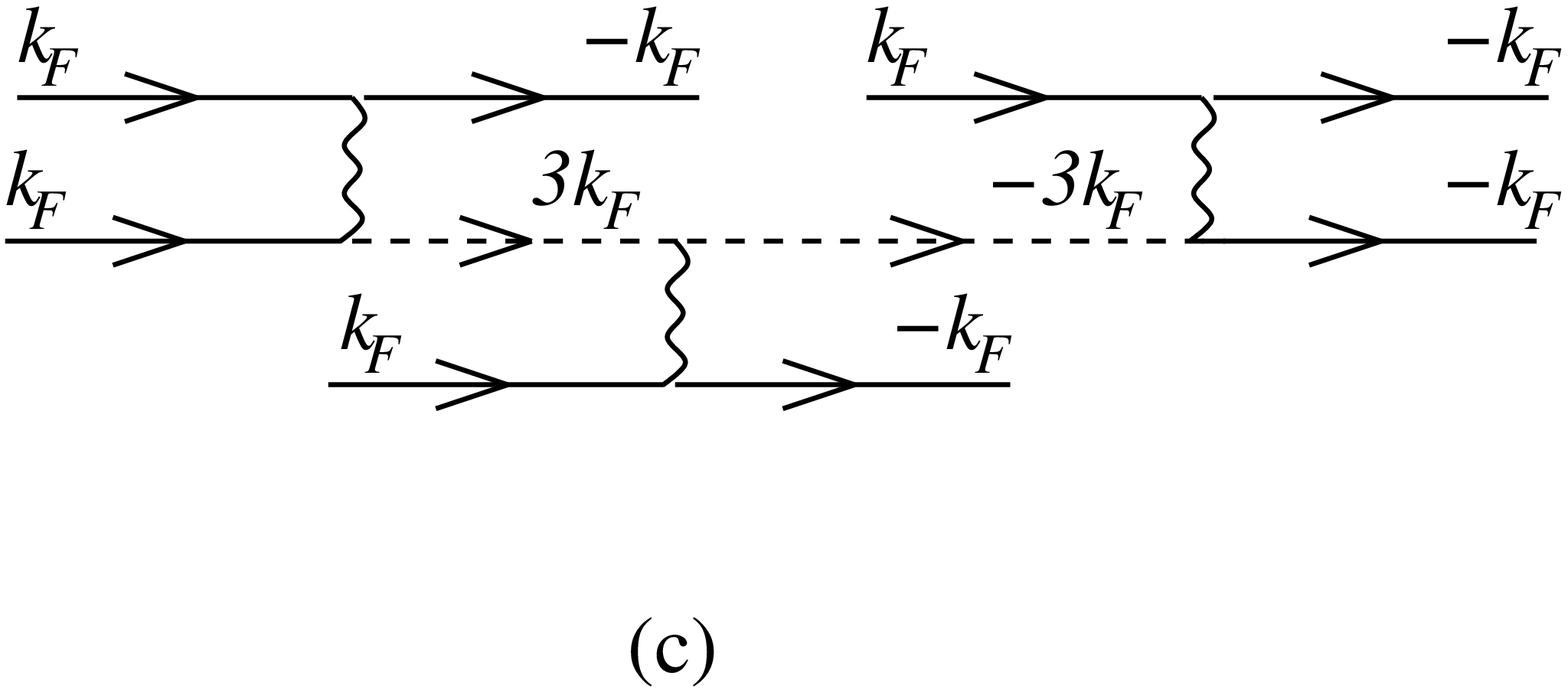}
    \caption{The diagrams that give rise to the $\cos 4\phi_{\rho+}$ term.}
    \label{fig:umklapps}
  \end{center}
\end{figure}

In this section, we give a derivation of the $\cos 4\phi_{\rho+}$ term
in the bosonized charge Hamiltonian (\ref{eq:charge-hamiltonian}) of the
quarter filled ladder following Ref.~\onlinecite{tsuchiizu_qf1d}.
To derive this term, one needs to separate the
 low energy processes which keep all the particles
near the Fermi points $\pm k_F$ from the high energy
 processes that involve transfer of particles near the points $\pm
 3k_F$.  By eliminating the latter high energy processes, one is
left with an effective action that involve only the Fermi points.
The details of
the procedure to eliminate the high energy states
are exposed in   Ref.~\onlinecite{tsuchiizu_qf1d}.
The processes that give rise to the $\cos 4\phi_{\rho+}$ are represented
diagrammatically on figure~\ref{fig:umklapps}.

The corresponding terms in the action read for diagram (a):
\begin{eqnarray}
  \label{eq:action-a}
  S_a &= & \frac 1 {2L^3} \sum_{r=R,L} \sum_{k_1,k'_1\atop q_1}
  \sum_{p_1,\sigma_1 \atop p'_1,\sigma'_1}
  \int_0^\beta d\tau_1\sum_{k_2,k'_2\atop q_2} \sum_{p_2,\sigma_2
    \atop p'_2,\sigma'_2}
  \int_0^\beta d\tau_2 \sum_{k_3,k'_3\atop q_3} \sum_{p_3,\sigma_3 \atop p'_3,\sigma'_3}
  \int_0^\beta d\tau_3 X_{\sigma_1,\sigma'_1}^{p_1,p'_1}
  X_{\sigma_2,\sigma'_2}^{p_2,p'_2}
  Y_{\sigma_3,\sigma'_3}^{p_3,p'_3}\times \nonumber \\
& &\langle (c^\dagger_{k_1+q_1,-r,p_1,\sigma_1}
c^\dagger_{k'_1-q_1,-r,p'_1,\sigma'_1}c_{k'_1,r,p'_1,\sigma'_1}
d_{k_1,-r,p_1,\sigma_1})(\tau_1)
(c^\dagger_{k_2+q_2,-r,p_2,\sigma_2}c^\dagger_{k'_2-q_2,-r,p'_2,\sigma'_2}
c_{k'_2,r,p'_2,\sigma'_2}
d_{k_2,-r,p_2,\sigma_2})(\tau_2)\times \nonumber \\
& & (d^\dagger_{k_3+q_3,-r,p_3,\sigma_3}d^\dagger_{k'_3-q_3,-r,p'_3,\sigma'_3}
c_{k'_3,r,p'_3,\sigma'_3}c_{k_3,r,p_3,\sigma_3})(\tau_3) \rangle_d,
\end{eqnarray}
\noindent for diagram (b):
\begin{eqnarray}
  \label{eq:action-b}
  S_b &= & \frac 1 {2L^3} \sum_{r=R,L} \sum_{k_1,k'_1\atop q_1}
  \sum_{p_1,\sigma_1 \atop p'_1,\sigma'_1}
  \int_0^\beta d\tau_1\sum_{k_2,k'_2\atop q_2} \sum_{p_2,\sigma_2
    \atop p'_2,\sigma'_2}
  \int_0^\beta d\tau_2 \sum_{k_3,k'_3\atop q_3} \sum_{p_3,\sigma_3 \atop p'_3,\sigma'_3}
  \int_0^\beta d\tau_3 X_{\sigma_1,\sigma'_1}^{p_1,p'_1}
  X_{\sigma_2,\sigma'_2}^{p_2,p'_2}
  Z_{\sigma_3,\sigma'_3}^{p_3,p'_3}\times \nonumber \\
& & \langle (d^\dagger_{k_1+q_1,r,p_1,\sigma_1}
c^\dagger_{k'_1-q_1,-r,p'_1,\sigma'_1} c_{k'_1,r,p'_1,\sigma'_1}
c_{k_1,r,p_1,\sigma_1})(\tau_1) (c^\dagger_{k_2+q_2,-r,p_2,\sigma_2}
c^\dagger_{k'_2-q_2,-r,p'_2,\sigma'_2} c_{k'_2,r,p'_2,\sigma'_2}
d_{k_2,-r,p_2,\sigma_2})(\tau_2)\times \nonumber \\
& & (c^\dagger_{k_3+q_3,-r,p_3,\sigma_3}
d^\dagger_{k'_3-q_3,-r,p'_3,\sigma'_3} c_{k'_3,r,p'_3,\sigma'_3}
d_{k_3,r,p_3,\sigma_3})(\tau_3)\rangle_d,
\end{eqnarray}
\noindent for diagram (c):
\begin{eqnarray}
  \label{eq:action-c}
  S_c &= & \frac 1 {2L^3} \sum_{r=R,L} \sum_{k_1,k'_1\atop q_1}
  \sum_{p_1,\sigma_1 \atop p'_1,\sigma'_1}
  \int_0^\beta d\tau_1\sum_{k_2,k'_2\atop q_2} \sum_{p_2,\sigma_2
    \atop p'_2,\sigma'_2}
  \int_0^\beta d\tau_2 \sum_{k_3,k'_3\atop q_3} \sum_{p_3,\sigma_3 \atop p'_3,\sigma'_3}
  \int_0^\beta d\tau_3 X_{\sigma_1,\sigma'_1}^{p_1,p'_1}
  X_{\sigma_2,\sigma'_2}^{p_2,p'_2}
  X_{\sigma_3,\sigma'_3}^{p_3,p'_3}\times \nonumber \\
& & \langle (d^\dagger_{k_1+q_1,r,p_1,\sigma_1}
c^\dagger_{k'_1-q_1,-r,p'_1,\sigma'_1} c_{k'_1,r,p'_1,\sigma'_1}c_{
  k_1,r,p_1,\sigma_1})(\tau_1) (c^\dagger_{k_2+q_2,-r,p_2,\sigma_2}
c^\dagger_{k'_2-q_2,-r,p'_2,\sigma'_2}
c_{k'_2,r,p'_2,\sigma'_2}d_{k_2,-r,p_2,\sigma_2})(\tau_2) \times
\nonumber \\
& &(c^\dagger_{k_3+q_3,-r,p_3,\sigma_3}
d^\dagger_{k'_3-q_3,-r,p'_3,\sigma'_3} d_{k'_3,-r,p'_3,\sigma'_3}
c_{k_3,r,p_3,\sigma_3})(\tau_3)\rangle_d,
\end{eqnarray}
\noindent where we use notations similar to those of
Ref.~\onlinecite{tsuchiizu_qf1d}. The differences with
Ref.~\onlinecite{tsuchiizu_qf1d} are the following. The most important
is the apparition of a chain index $p=1,2$ for the fermions. Another
difference (very minor) is that we write the annihilation operator
for fermions with momentum close to $k_F$, $c_R$ and $d_R$ for
fermions with momentum close to $3k_F$. For fermions with momentum
close to $-k_F$ and $-3k_F$ we write the respective annihilation
operators $c_L$ and $d_L$. We also have by definition $-R=L$ and
$-L=R$. We have the following expressions for $X,Y,Z$:
\begin{eqnarray}
  \label{eq:XYZ-UV}
  X_{\sigma \sigma'}^{p p'}&=& \frac {U a}2
    \delta_{\sigma,-\sigma'}\delta_{p,p'} + V_\perp a
    (1-\delta_{p,p'}) \\
  Y_{\sigma \sigma'}^{p p'}&=&(Ua\delta_{\sigma,-\sigma'} -2 V_\parallel)\delta_{p,p'} + V_\perp a
    (1-\delta_{p,p'}) \\
  Z_{ \sigma \sigma'}^{p p'}&=&( \frac {U a}2
    \delta_{\sigma,-\sigma'}-  V_\parallel)\delta_{p,p'} + V_\perp a
    (1-\delta_{p,p'})
\end{eqnarray}

The operators $d_R,d_L$ annihilate states of high energy. In
the low energy limit, these states do not appear and should therefore
be integrated out\cite{tsuchiizu_qf1d}.
This is the meaning of the symbol $\langle \ldots
\rangle_d$. These states are integrated out by using the following
Green's function for the $d$ fermions:
\begin{eqnarray}
  \label{eq:Green-d}
  \langle d_{k,r,p,\sigma} (\tau) d_{k',r',p',\sigma'}(0)\rangle =
  -\frac 1 {2\sqrt{2} t} \delta_{k,k'} \delta_{r,r'} \delta_{p,p'}
  \delta_{\sigma,\sigma'}
\end{eqnarray}
where $t$ is the transfer integral of the extended Hubbard model.
This integration gives rise to a term that describes the
collision of four fermions with $8k_F$ Umklapp. This term reads:
\begin{eqnarray}
  \label{eq:low-energy-action}
  S=\frac 1 {8t^2} \int dx A_{\sigma_1 \sigma'_1 \sigma_2
    \sigma'_2}^{p_1 p'_1 p_2 p'_2}
  (\psi^\dagger_{R,p_1,\sigma_1} \psi_{L,p_1,\sigma_1})
(\psi^\dagger_{R,p'_1,\sigma'_1} \psi_{L,p'_1,\sigma'_1})
(\psi^\dagger_{R,p_2,\sigma_2} \psi_{L,p_2,\sigma_2})
(\psi^\dagger_{R,p'_2,\sigma'_2} \psi_{L,p'_2,\sigma'_2})
+\text{H. c.} .
\end{eqnarray}

\noindent If all the fermions
belong to the same chain($p_1=p'_1=p_2=p'_2$) , one recovers the term $\cos
4\sqrt{2}\phi_{\rho,p}$ derived in Ref.~\onlinecite{tsuchiizu_qf1d}.
For the case $(p_1,\sigma_1)=(1,\uparrow)$,
$(p'_1,\sigma'_1)=(1,\downarrow)$   $(p_2,\sigma_2)=(2,\uparrow)$,
$(p'_2,\sigma'_2)=(2,\downarrow)$ (and all the $4!=24$ cases
equivalent by permutation), the Umklapp term
(\ref{eq:low-energy-action}) can be bosonized in the
form:
\begin{eqnarray}
  \label{eq:umklapp-2ch}
  S=g \int dx
  \cos(2(\phi_{1,\uparrow}+\phi_{1,\downarrow}+\phi_{2,\uparrow}+\phi_{2,\downarrow})) = g \int \cos 4\phi_{\rho+}
\end{eqnarray}
This leads to the term we derived phenomenologically.
Finding the expression of $g$ is only a tedious calculation. The final
result is:
\begin{eqnarray}
  \label{eq:expression-g}
  g=\frac{V_\perp(U^2+10 V_\perp U + 4V_\perp^2-4V_\parallel
    V_\perp)\alpha}{32 t^2 \pi^2\alpha^2}
\end{eqnarray}

\section{Phenomenological spin density}\label{app:phen-spin-dens}

In this appendix, we give a phenomenological derivation of
 the $2n k_F$ Fourier components of the spin density that generalizes
 the equations~(\ref{eq:Sx})--(\ref{eq:Sz}). Let us begin with
 $S^z(x)$.
If we have a system of fermions with both spins up and down, we can
write the following expression\cite{haldane_bosons} for the spin density:
\begin{equation}
  \label{eq:density-phen}
  \rho_\alpha(x)= \rho_0 -\frac 1 \pi \partial_x \phi_\alpha + \sum_m
  A_m \cos 2 m ( \phi_\alpha -k_F x)
\end{equation}
Expressing these quantities in terms of $\phi_\rho,\phi_\sigma$,
forming the difference and taking into account the term $\cos
\sqrt{8}\phi_\sigma$ in the Hamiltonian, we obtain the expression:
\begin{equation}
  \label{eq:S-z-phen-0}
  S^z(x)=-\frac{1}{\pi \sqrt{2}} \partial_x \phi_\sigma + \sum_m  B_m \sin
  \sqrt{2} m \phi_\sigma \sin (m \sqrt{2}\phi_\rho -2 k_F x).
\end{equation}
Since the Hamiltonian of spin excitations contains a term $\cos
\sqrt{8}\phi_\sigma$, new terms will be generated by fusion of this
operator with $S^z(x)$. For $m$ odd, the fusions will generate a term
$\sin \sqrt{2}\phi_{\sigma}$. For $m$ even, the fusions will generate
a term $\partial_x \phi_{\sigma}$. This leads to the following
phenomenological expression for $S^z(x)$:
 \begin{eqnarray}
  \label{eq:S-z-phen}
  S^z(x)&=&-\frac{1}{\pi \sqrt{2}} \partial_x \phi_\sigma \sum_m
  A_{2m,z} \sin 2m (\sqrt{2}\phi_\rho -2 k_F x) \nonumber \\
&& + \sum_m  A_{2m+1,z} \sin
  \sqrt{2} \phi_\sigma \sin (2m+1)(\sqrt{2}\phi_\rho -2 k_F x).
\end{eqnarray}
The usual expression of the spin density~(\ref{eq:Sz}) is recovered for $A_{2m,z}=0$
and $A_{2m+1,z}=0$ for $m\ge 1$.
For the $S^{x,y}(x)$ similar expressions can be obtained.
Starting from the phenomenological Haldane expansion of the fermion
creation and annihilation operators\cite{haldane_bosons}:
\begin{equation}
  \label{eq:fermion-haldane}
  \psi_{\sigma}(x) \sim e^{\frac{i}{\sqrt{2}}(\theta_\rho+\sigma
    \theta_\sigma)} \sum_{m=-\infty}^{\infty}  e^{i(2m+1)
    [\frac{(\phi_\rho+\sigma \phi_{\sigma})}{\sqrt{2}} -k_F x]},
\end{equation}
\noindent we easily obtain:
\begin{eqnarray}
  \label{eq:Splus-phen-0}
  S^+(x)\sim e^{i\sqrt{2} \theta_\sigma} \sum_{m,m'} e^{i[
    (m-m')(\sqrt{2}\phi_\rho-2k_F x) + (m+m'+1) \sqrt{2} \phi_\sigma]}.
\end{eqnarray}
\noindent We note that $m+m'+1$ and $m-m'$ have different parities. We
see easily that by fusion with $\cos \sqrt{8} \phi_{\sigma}$, the
terms in $\phi_{\sigma}$ with $m+m'+1$ odd will be reduced to $\cos
\sqrt{2}\phi_{\sigma}$ while the terms with $m+m'+1$ even will be
reduced to $1$. The expression of $S^+(x)$ therefore reduces to:
\begin{eqnarray}
  \label{eq:Splus-phen}
  S^+(x)&\sim& e^{i\sqrt{2} \theta_\sigma} \sum_{m} A_{2m+1,x}
  e^{i(2m+1)(\sqrt{2}\phi_\rho-2k_F x)} \nonumber \\
        && +  e^{i\sqrt{2} \theta_\sigma} \cos \sqrt{2} \phi_{\sigma} \sum_{m} A_{2m,x}
  e^{i 2m(\sqrt{2}\phi_\rho-2k_F x)}.
\end{eqnarray}
We can check that the expressions we have obtained lead to rotational
invariant expression of the spin correlations for $K_\sigma=1$,
provided that $A_{m,z}=A_{m,x}$ for all $m$.

Eqs.~(\ref{eq:S-z-phen})-(\ref{eq:Splus-phen}) can also be derived
from a more physical argument. The Ogata-Shiba
wavefunction tell us that in the limit of $U/t\gg 1$, the
spin excitations
can be described as a ``squeezed'' antiferromagnetic spin chains,
the spins being carried by the electrons\cite{ogata_inf}.
The spin density should therefore be described by the following
expression:
\begin{eqnarray}
  \label{eq:spin-particle}
  \vec{S}(x)=\sum_n \vec{S}_n \delta(x-x_n),
\end{eqnarray}
\noindent where the $x_i$ are the positions of the electrons along the
chain, labelled in such way that $x_1<x_2<\ldots<x_N$.
The vector $\vec{S}_n$ can be decomposed into a staggered and a
uniform component  as $\vec{S}_n=\vec{J}_n + (-)^n \vec{n}_n$. We will
assume that both $\vec{J}_n$ and $\vec{n}_n$ are slowly varying at the
scale of the average interparticle distance so that we
can write:  $\vec{J}_n=\vec{J}(x=x_n)$, $\vec{n}_n=\vec{n}(x=x_n)$,
the functions $\vec{J}(x)$, $\vec{n}(x)$ varying smoothly between the
points $x_n$. We note that the integral from $-\infty$ to $+\infty$ of
$\vec{J}$ is the total magnetization operator and also the generator
of the rotations. Thus, we expect the functions $\vec{J}$ and
$\vec{n}$  to obey the following commutation relations:
\begin{eqnarray}
  \label{eq:commutation-Jn}
  \lbrack J^a(x),J^b(x')\rbrack&=&i\epsilon_{abc} J^c(x)\delta(x-x') (a\ne b), \\
  \lbrack J^a(x),n^b(x')\rbrack&=&i\epsilon_{abc} n^c(x)\delta(x-x') (a\ne b),
\end{eqnarray}
\noindent that coincide with the usual commutation relation of the
generator of the rotations and the staggered magnetic
field\cite{affleck_houches}. Moreover, since the spin excitations of a
spinful Luttinger liquid are expected to be described by a single
gapless boson, as those of a spin-1/2 chain,
it is natural to identify $\vec{J}$ to the bosonized
uniform spin density of a spin 1/2 chain, and $\vec{n}$ to the
staggered spin density. The corresponding expression reads:
\begin{eqnarray}
  \label{eq:spinchain}
  J^+(x)=J^x+iJ^y\sim e^{i\sqrt{2}\theta_\sigma} \cos \sqrt{2}\phi_\sigma
  ; n^+ \sim e^{i\sqrt{2}\theta_\sigma} \nonumber \\
  J^z=-\frac{1}{\pi \sqrt{2}}\phi_\sigma ; n^z \sim \sin \sqrt{2}\phi_\sigma
\end{eqnarray}
\noindent
Let us now introduce a function $\phi(x)$ such that $\phi(x_n)=n\pi$.
We can rewrite the delta function as:
\begin{eqnarray}
  \sum_n \delta(x-x_n)=\sum_n \delta(\phi(x)-n\pi)\frac{d\phi}{dx} =
  \sum_m e^{i 2m\phi(x)}\frac{d\phi}{dx},
\end{eqnarray}
\noindent and we have: $e^{i\phi(x)}=(-)^n$. This allows us to write:
\begin{eqnarray}
  \label{eq:spin-pheno}
  \vec{S}(x)= \vec{J}(x) \sum_m e^{i 2m\phi(x)}\frac{d\phi}{dx}
  +\vec{n}(x)  \sum_m e^{i (2m+1)\phi(x)}\frac{d\phi}{dx}.
\end{eqnarray}
Since $\phi(x)$ must be $n\pi$ each time that there is a particle, we have
that $\phi(x)=\phi_\uparrow(x)+\phi_\downarrow(x)$, from which
we easily obtain: $\phi(x)=\pi \rho_0 -\sqrt{2}\phi_\rho$. Using the
bosonized expressions~(\ref{eq:spinchain})
of $\vec{J}$ and $\vec{n}$ in terms of
$\phi_{\sigma}$ the
expressions (\ref{eq:S-z-phen})-~(\ref{eq:Splus-phen}) are then easily
seen to be equivalent to (\ref{eq:spin-pheno}).
Applying the expression (\ref{eq:spin-pheno}) to our problem we see
that the terms coming from the $4k_F$ component of the
spin density are less relevant and read:
\begin{equation}
  (\cos 4\phi_{\rho+} + \cos 4\phi_{\rho-}) \vec{J}_1\cdot \vec{J}_2
\end{equation}
The contribution of these terms to the Hamiltonian (\ref{eq:full}) can
thus usually be neglected being less relevant. However, in the case of
$J_\perp$ sufficiently large, a gap may be formed in the modes
$\rho-,\sigma\pm$ at higher energy scale than in $\rho+$. In that case, the
terms we have derived can lead to a modification of the coefficient
$g_0$  of
the term $\cos 4\phi_{\rho+}$ in (\ref{eq:charge-hamiltonian}) and a
change of the ground state expectation value of $\phi_{\rho+}$ from
$\frac{\pi}{4}$ to $0$ if $J_\perp$ is large enough.


\begin{thebibliography}{100}

\bibitem{cheong_mang}
C. Chen and S. Cheong, Phys. Rev. Lett. {\bf 76},  4042  (1996).

\bibitem{mori_mang}
S. Mori, C. Chen, and S. Cheong, Nature {\bf 392},  473  (1998).

\bibitem{monceau_co_1d}
P. Monceau, F. Nad, and S. Brazovskii, Phys. Rev. Lett. {\bf 86},  4080
  (2001).

\bibitem{tranquada_nick}
J.~M. Tranquada, D.~J. Buttrey, and V. Sachan, Phys. Rev. B {\bf 54},  12318
  (1996).

\bibitem{tranquada_cup}
N. Ichikawa {\it et~al.}, Phys. Rev. Lett. {\bf 85},  1738  (2000).

\bibitem{wigner}
E. Wigner, Trans. Faraday Soc. {\bf 34},  678  (1938).

\bibitem{emery_revue_1d}
V.~J. Emery,  in {\em Highly Conducting One-Dimensional Solids}, edited by
  J.~T. Devreese, R.~P. Evrard, and V.~E. van Doren (Plenum Press, New York and
  London, 1979).

\bibitem{solyom_revue_1d}
J. S{\'o}lyom, Adv. Phys. {\bf 28},  209  (1979).

\bibitem{schulz_houches_revue}
H.~J. Schulz,  in {\em Mesoscopic quantum physics, Les Houches LXI}, edited by
  E. Akkermans, G. Montambaux, J.~L. Pichard, and J. Zinn-Justin (Elsevier,
  Amsterdam, 1995), p.\ 533.

\bibitem{white_dmrg_letter}
S.~R. White, Phys. Rev. Lett. {\bf 69},  2863  (1992).

\bibitem{hubbard_wigner}
J. Hubbard, Phys. Rev. B {\bf 17},  494  (1978).

\bibitem{schulz_mott_revue}
H.~J. Schulz,  in {\em ``Strongly correlated electronic materials'' The Los
  Alamos symposium 1993}, edited by K.~S. Bedell {\it et~al.} (Addison-Westley,
  Reading, Massachusetts, 1994).

\bibitem{giamarchi_mott_shortrev}
T. Giamarchi, Physica B {\bf 230-232},  975  (1997).

\bibitem{tsuchiizu_qf1d}
M. Tsuchiizu, H. Yoshioka, and Y. Suzumura, J. Phys. Soc. Jpn. {\bf 70},  1460
  (2001).

\bibitem{poilblanc_2ch}
D. Poilblanc, H. Tsunetsugu, and T.~M. Rice, Phys. Rev. B {\bf 50},  6511
  (1994).

\bibitem{white_stripes_tj}
S.~R. White and D.~J. Scalapino, Phys. Rev. Lett. {\bf 80},  1272  (1998).

\bibitem{bonca_3leg}
J. Bonca {\it et~al.}, Phys. Rev. B {\bf 61},  3261  (2000).

\bibitem{tohyama_4leg}
T. Tohyama {\it et~al.}, Phys. Rev. B {\bf 59},  R11649  (1999).

\bibitem{white_ladder_friedel}
S.~R. White, I. Affleck, and D.~J. Scalapino, Phys. Rev. B {\bf 65},  165122
  (2002).

\bibitem{haldane_gap}
F.~D.~M. Haldane, Phys. Rev. Lett. {\bf 50},  1153  (1983).

\bibitem{takano_spingap}
M. Takano {\it et~al.}, Phys. Rev. Lett. {\bf 73},  3463  (1994).

\bibitem{chaboussant_cuhpcl}
G. Chaboussant {\it et~al.}, Phys. Rev. B {\bf 55},  3046  (1997).

\bibitem{iwase_cav2o5}
H. Iwase, M. Isobe, Y. Ueda, and H. Yasuoka, J. Phys. Soc. Jpn. {\bf 65},  2397
   (1996).

\bibitem{rovira_dtttf}
C. Rovira {\it et~al.}, Angew. Chem. Inter. Ed. Engl. {\bf 36},  2324  (1997).

\bibitem{watson_bpcb}
B.~C. Watson {\it et~al.}, Phys. Rev. Lett. {\bf 86},  5168  (2001).

\bibitem{landee_ladder}
C.~P. Landee {\it et~al.}, Phys. Rev. B {\bf 63},  100402(R)  (2001).

\bibitem{rice_srcuo}
T. Rice, S. Gopalan, and M. Sigrist, Europhys. Lett. {\bf 23},  445  (1993).

\bibitem{fabrizio_2ch_rg}
M. Fabrizio, Phys. Rev. B {\bf 48},  15838  (1993).

\bibitem{nagaosa_2ch}
N. Nagaosa, Sol. State Comm. {\bf 94},  495  (1995).

\bibitem{schulz_2chains}
H.~J. Schulz, Phys. Rev. B {\bf 53},  R2959  (1996).

\bibitem{balents_2ch}
L. Balents and M.~P.~A. Fisher, Phys. Rev. B {\bf 53},  12133  (1996).

\bibitem{fabrizio_q1d}
M. Fabrizio, A. Parola, and E. Tosatti, Phys. Rev. B {\bf 46},  3159  (1992).

\bibitem{poilblanc_2ch_mc}
D. Poilblanc, D.~J. Scalapino, and W. Hanke, Phys. Rev. B {\bf 52},  6796
  (1995).

\bibitem{troyer_2ch_tj}
M. Troyer, H. Tsunetsugu, and T.~M. Rice, Phys. Rev. B {\bf 53},  251  (1996).

\bibitem{hayward_2chain_2}
C.~A. Hayward and D. Poilblanc, Phys. Rev. B {\bf 53},  11721  (1996).

\bibitem{noack_dmrg_2ch}
R. Noack, S. White, and D. Scalapino, Phys. Rev. Lett. {\bf 73},  882  (1994).

\bibitem{uchara_SrCaCuO}
M. Uchara {\it et~al.}, J. Phys. Soc. Jpn. {\bf 65},  2764  (1996).

\bibitem{isobe_nav2o5}
M. Isobe and Y. Ueda, J. Phys. Soc. Jpn. {\bf 65},  1178  (1996).

\bibitem{boer}
A. Meetsma {\it et~al.}, Acta Crystallogr. Sect. C {\bf 54},  1558  (1998).

\bibitem{smolinski_nav2o5}
H. Smolinski {\it et~al.}, Phys. Rev. Lett. {\bf 80},  5164  (1998).

\bibitem{mostovoy_nav2o5}
M. Mostovoy and D. Khomskii, Sol. State Comm. {\bf 113},  159  (1999), cond-mat
  9806215.

\bibitem{seo_nav2o5}
H. Seo and H. Fukuyama, J. Phys. Soc. Jpn. {\bf 67},  2602  (1998).

\bibitem{ohama_nav2o5}
T. Ohama, H. Yasuoka, M. Isobe, and Y. Ueda, Phys. Rev. B {\bf 59},  3299
  (1999).

\bibitem{fagot-revurat_nav2o5}
Y. Fagot-Revurat, M. Mehring, and R.~K. Kremer, Phys. Rev. Lett. {\bf 84},
  4176  (2000).

\bibitem{grenier_nav2o5}
B. Grenier {\it et~al.}, Phys. Rev. Lett. {\bf 86},  5966  (2001).

\bibitem{sawa_nav2o5}
H. Sawa {\it et~al.}, J. Phys. Soc. Jpn. {\bf 71},  385  (2002).

\bibitem{vojta_qfl_short}
M. Vojta, R.~E. Hetzel, and R.~M. Noack, Phys. Rev. B {\bf 60},  R4817  (1999).

\bibitem{vojta_qfl}
M. Vojta, A. Hubsch, and R. Noack, Phys. Rev. B {\bf 63},  045105  (2001).

\bibitem{lin_so8}
H. Lin, L. Balents, and M.~P.~A. Fisher, Phys. Rev. B {\bf 58},  1794  (1998).

\bibitem{konik_ff_so8}
R. Konik and A.~W.~W. Ludwig, Phys. Rev. B {\bf 64},  155112  (2001).

\bibitem{wu_2leg_firstorder}
C. Wu, W.~V. Liu, and E. Fradkin, cond-mat/0206248 (unpublished).

\bibitem{tsuchiizu_2leg_firstorder}
M. Tsuchiizu and A. Furusaki, Phys. Rev. B {\bf 66},  245106  (2002).

\bibitem{amasaki_2ch_zigzag}
R. Amasaki, Y. Shibata, and Y. Ohta, Phys. Rev. B {\bf 66},  012502  (2002).

\bibitem{horsch_nav2o5}
P. Horsch and F. Mack, Eur. Phys. J. B {\bf 5},  367  (1998).

\bibitem{sa_nav2o5}
D. Sa and C. Gros, Eur. Phys. J. B {\bf 18},  421  (2000).

\bibitem{brazovskii_transhop}
S. Brazovskii and V. Yakovenko, J. de Phys. (Paris) Lett. {\bf 46},  L111
  (1985).

\bibitem{tsunegutsu_2ch}
H. Tsunetsugu, M. Troyer, and T.~M. Rice, Phys. Rev. B {\bf 49},  16078
  (1994).

\bibitem{ogata_inf}
M. Ogata and H. Shiba, Phys. Rev. B {\bf 41},  2326  (1990).

\bibitem{haldane_xxzchain}
F.~D.~M. Haldane, Phys. Rev. Lett. {\bf 45},  1358  (1980).

\bibitem{jeckelmann_hubbard1d}
E. Jeckelmann, F. Gebhard, and F.~H.~L. Essler, Phys. Rev. Lett. {\bf 85},
  3910  (2000).

\bibitem{controzzi_mott}
D. Controzzi, F.~H.~L. Essler, and A.~M. Tsvelik,  in {\em New Theoretical
  approaches to strongly correlated systems}, Vol.~23 of {\em NATO Science
  Series II. Mathematics, Physics and Chemistry}, edited by A.~M. Tsvelik
  (Kluwer Academic Publishers, Dordrecht, 2001), p.\ 25.

\bibitem{shankar_spinless_conductivite}
R. Shankar, Int. J. Mod. Phys. B {\bf 4},  2371  (1990).

\bibitem{black_equ}
J.~L. Black and V.~J. Emery, Phys. Rev. B {\bf 23},  429  (1981).

\bibitem{nijs_equivalence}
M.~P.~M. den Nijs, Phys. Rev. B {\bf 23},  6111  (1981).

\bibitem{schulz_hubbard_exact}
H.~J. Schulz, Phys. Rev. Lett. {\bf 64},  2831  (1990).

\bibitem{kawakami_tj}
N. Kawakami and S.~K. Yang, Phys. Rev. Lett. {\bf 65},  2309  (1990).

\bibitem{kawakami_hubbard}
N. Kawakami and S.~K. Yang, Phys. Lett. A {\bf 148},  359  (1990).

\bibitem{frahm_confinv}
H. Frahm and V.~E. Korepin, Phys. Rev. B {\bf 42},  10553  (1990).

\bibitem{mila_zotos}
F. Mila and X. Zotos, Europhys. Lett. {\bf 24},  133  (1993).

\bibitem{sano_extended_hubbard_1d}
K. Sano and Y. {\=O}no, J. Phys. Soc. Jpn. {\bf 63},  1250  (1994).

\bibitem{nakamura_tJ}
M. Nakamura, K. Nomura, and A. Kitazawa, Phys. Rev. Lett. {\bf 79},  3214
  (1997), cond-mat/9708204.

\bibitem{haldane_bosons}
F.~D.~M. Haldane, Phys. Rev. Lett. {\bf 47},  1840  (1981).

\bibitem{schulz_wigner_1d}
H.~J. Schulz, Phys. Rev. Lett. {\bf 71},  1864  (1993).

\bibitem{khveshenko_2chain}
D.~V. Khveshenko and T.~M. Rice, Phys. Rev. B {\bf 50},  252  (1994).

\bibitem{oshikawa_plateaus}
M. Oshikawa, M. Yamanaka, and I. Affleck, Phys. Rev. Lett. {\bf 78},  1984
  (1997).

\bibitem{cardy_scaling}
J.~L. Cardy, {\em Scaling and Renormalization in Statistical Physics}, {\em
  Cambridge Lecture Notes in Physics} (Cambridge University Press, Cambdridge,
  UK, 1996).

\bibitem{riera_coexistence_1d}
J. Riera and D. Poilblanc, Phys. Rev. B {\bf 62},  R16243  (2000).

\bibitem{essler_mott_excitons1d}
F.~H.~L. Essler, F. Gebhard, and E. Jeckelmann, Phys. Rev. B {\bf 64},  125119
  (2001).

\bibitem{white_zigzag}
S.~R. White and I. Affleck, Phys. Rev. B {\bf 54},  9862  (1996).

\bibitem{allen}
D. Allen and D. S{\'e}n{\'e}chal, Phys. Rev. B {\bf 55},  299  (1997).

\bibitem{allen_spinons}
D. Allen, F.~H.~L. Essler, and A.~A. Nersesyan, Phys. Rev. B {\bf 61},  8871
  (2000).

\bibitem{witten_shankar}
E. Witten and R. Shankar, Nucl. Phys. B {\bf 141},  349  (1978).

\bibitem{zachar_exotic_kondo}
O. Zachar and A.~M. Tsvelik, Phys. Rev. B {\bf 63},  033103  (2001),
  cond-mat/9909296.

\bibitem{fujimoto_kondo1d}
S. Fujimoto and N. Kawakami, J. Phys. Soc. Jpn. {\bf 63},  4322  (1994).

\bibitem{lehur98_kondo1d}
K. {Le Hur}, Phys. Rev. B {\bf 58},  10261  (1998).

\bibitem{lehur00_kondo1d}
K. {Le Hur}, Phys. Rev. B {\bf 62},  4408  (2000).

\bibitem{schulz_son}
H. Schulz, cond-mat/9808167 (unpublished).

\bibitem{azaria_su4}
P. Azaria, A.~O. Gogolin, P. Lecheminant, and A.~A. Nersesyan, Phys. Rev. Lett.
  {\bf 83},  624  (1999).

\bibitem{shelton_spin_ladders}
D.~G. Shelton, A.~A. Nersesyan, and A.~M. Tsvelik, Phys. Rev. B {\bf 53},  8521
   (1996).

\bibitem{assaraf_su(n)}
R. Assaraf, P. Azaria, M. Caffarel, and P. Lecheminant, Phys. Rev. B {\bf 60},
  2299  (1999), cond-mat/9903057.

\bibitem{azaria_su4_long}
P. Azaria, E. Boulat, and P. Lecheminant, Phys. Rev. B {\bf 61},  12112
  (1999).

\bibitem{itoi_spin_orbital}
C. Itoi, S. Qin, and I. Affleck, Phys. Rev. B {\bf 61},  6747  (2000),
  cond-mat/9910109.

\bibitem{pati_orbital_dmrg}
S.~K. Pati, R.~R.~P. Singh, and D.~I. Khomskii, Phys. Rev. Lett. {\bf 81},
  5406  (1998).

\bibitem{nersesyan_biquadratic}
A. Nersesyan and A.~M. Tsvelik, Phys. Rev. Lett. {\bf 78},  3939  (1997), ibid.
  , \textbf{79}, E 1171.

\bibitem{orignac_spintube}
E. Orignac, R. Citro, and N. Andrei, Phys. Rev. B {\bf 61},  11533  (2000).

\bibitem{fabrizio_dsg}
M. Fabrizio, A.~O. Gogolin, and A.~A. Nersesyan, Nucl. Phys. B {\bf 580},  647
  (2000).

\bibitem{shankar_gn_at}
R. Shankar, Phys. Rev. Lett. {\bf 55},  453  (1985).

\bibitem{goldschmidt_susy}
Y.~Y. Goldschmidt, Phys. Rev. Lett. {\bf 56},  1627  (1986).

\bibitem{japaridze_cic_transition}
G.~I. Japaridze and A.~A. Nersesyan, JETP Lett. {\bf 27},  334  (1978).

\bibitem{pokrovsky_talapov_prl}
V.~L. Pokrovsky and A.~L. Talapov, Phys. Rev. Lett. {\bf 42},  65  (1979).

\bibitem{schulz_cic2d}
H.~J. Schulz, Phys. Rev. B {\bf 22},  5274  (1980).

\bibitem{schulz_mitwochain}
H.~J. Schulz, Phys. Rev. B {\bf 59},  R2471  (1999).

\bibitem{chitra_spinchains_field}
R. Chitra and T. Giamarchi, Phys. Rev. B {\bf 55},  5816  (1997).

\bibitem{affleck_houches}
I. Affleck,  in {\em Fields, Strings and Critical Phenomena}, edited by E.
  Brezin and J. Zinn-Justin (Elsevier Science Publishers, Amsterdam, 1988).

\end{thebibliography}

\end{document}